\documentclass[11pt]{article}
\pdfoutput=1
\usepackage{graphicx,rotating,hyperref,slashed,amsmath,xcolor,amssymb,amsfonts,colortbl,cite,amssymb}
\makeatletter
\hypersetup{colorlinks,bookmarksopen,bookmarksnumbered,
linkcolor=blus,pdfstartview=FitH,urlcolor=rossos,citecolor=verde}
\allowdisplaybreaks

\def\lsim{\mathrel{\rlap{\lower3pt\hbox{\hskip0pt$\sim$}}
   \raise1pt\hbox{$<$}}}         
\def\gsim{\mathrel{\rlap{\lower4pt\hbox{\hskip1pt$\sim$}}
   \raise1pt\hbox{$>$}}}         

 \newcommand{\sfootnote}[1]{}
\definecolor{bluc}{cmyk}{1,1,0,0.1}
\definecolor{rossoCP3}{cmyk}{0,.88,.77,.40}
\definecolor{rosso}{cmyk}{0,1,1,0.4}
\definecolor{rossos}{cmyk}{0,1,1,0.55}
\definecolor{rossoc}{cmyk}{0,1,1,0.2}
\definecolor{verdes}{cmyk}{0.92,0,0.59,0.4}

\hypersetup{colorlinks, bookmarksopen, bookmarksnumbered,
citecolor=verdes, linkcolor=bluc, pdfstartview=FitH, urlcolor=rossos}

\newcommand{\mio}[1]{}

\definecolor{Gray}{gray}{0.95}

\usepackage{multicol}
\usepackage{color}
\definecolor{rosso}{cmyk}{0,1,1,0.4}
\definecolor{rossos}{cmyk}{0,1,1,0.55}
\definecolor{rossoc}{cmyk}{0,1,1,0.2}
\definecolor{blu}{cmyk}{1,1,0,0.3}
\definecolor{blus}{cmyk}{1,1,0,0.6}
\definecolor{bluc}{cmyk}{1,1,0,0.1}
\definecolor{verde}{cmyk}{0.92,0,0.59,0.25}
\definecolor{verdec}{cmyk}{0.92,0,0.59,0.15}
\definecolor{verdes}{cmyk}{0.92,0,0.59,0.4}

\def\circa#1{\,\raise.3ex\hbox{$#1$\kern-.75em\lower1ex\hbox{$\sim$}}\,}

\newcommand{\beq}{\begin{equation}}
\newcommand{\eeq}{\end{equation}}

\newcommand{\bea}{\begin{eqnarray}}
\newcommand{\eea}{\end{eqnarray}}
\newcommand{\be}{\begin{equation}}
\newcommand{\ee}{\end{equation}}
\newfam\rsfsfam
\def\mathscr#1{{\fam\rsfsfam\relax#1}}

\def\circa#1{\,\raise.3ex\hbox{$#1$\kern-.75em\lower1ex\hbox{$\sim$}}\,}
\makeatletter

\def\hhref#1{\href{http://arxiv.org/abs/#1}{arXiv:#1}} 

\newcommand{\doi}[1]{\href{http://dx.doi.org/#1}{[doi]}}

\def\hhref#1{\href{http://arxiv.org/abs/#1}{arXiv:#1}} 
 
\def\art{\@ifnextchar[{\eart}{\oart}}
\def\eart[#1]#2#3#4#5#6{{\rm #2}, {\em #3 \bf #4} {\rm (#6) #5} ({\em #1})}

\def\article{\@ifnextchar[{\earticle}{\oarticle}}
\def\oarticle#1#2#3#4#5#6{{\rm #1}, {\em ``#6''}, {\rm #2 #3 (#5) #4}}
\def\earticle[#1]#2#3#4#5#6#7{{\rm #2}, {\em ``#7''}, {\rm #3 #4 (#6) #5}  [\hhref{#1}]}
\def\hepart[#1]#2{{\rm #2, \em#1}}
\def\heparticle[#1]#2#3{#2, {\em ``#3''} [\hhref{#1}]}

\newcounter{alphaequation}[equation]
\def\thealphaequation{\theequation\hbox to
0.6em{\hfil\alph{alphaequation}\hfil}}
\def\eqnsystem#1{
\def\@eqnnum{{\rm (\thealphaequation)}}
\def\@@eqncr{\let\@tempa\relax \ifcase\@eqcnt \def\@tempa{& & &} \or
  \def\@tempa{& &}\or \def\@tempa{&}\fi\@tempa
  \if@eqnsw\@eqnnum\refstepcounter{alphaequation}\fi
\global\@eqnswtrue\global\@eqcnt=0\cr}
\refstepcounter{equation} \let\@currentlabel\theequation \def\@tempb{#1}
\ifx\@tempb\empty\else\label{#1}\fi
\refstepcounter{alphaequation}
\let\@currentlabel\thealphaequation
\global\@eqnswtrue\global\@eqcnt=0 \tabskip\@centering\let\\=\@eqncr
$$\halign to \displaywidth\bgroup \@eqnsel\hskip\@centering
$\displaystyle\tabskip\z@{##}$&\global\@eqcnt\@ne
\hskip2\arraycolsep\hfil${##}$\hfil& \global\@eqcnt\tw@\hskip2\arraycolsep
$\displaystyle\tabskip\z@{##}$\hfil
\tabskip\@centering&\llap{##}\tabskip\z@\cr}
\def\endeqnsystem{\@@eqncr\egroup$$\global\@ignoretrue} \makeatother

\definecolor{fiorentina}{rgb}{.5,0,.5}

\begin{document}

\begin{center}
{\Large\bf{  A new scalar-tensor realization of  Ho\v{r}ava-Lifshitz gravity
}} 

 \vspace{0.9truecm}
\thispagestyle{empty} 

{  Javier Chagoya, Gianmassimo Tasinato}
\\[5mm]
{\it Department of Physics, Swansea University, Swansea, SA2 8PP, U.K.}
\end{center}

 \vspace{1.cm}
\begin{abstract}
\noindent
We discuss  a new  covariant scalar-tensor system aimed  to realise    
Ho\v{r}ava proposal for a power-counting  renormalizable  theory of gravity, with the special feature of  
   not propagating  scalar degrees of freedom in an appropriate gauge.  The theory is characterized by a new symmetry acting
  on the metric, that can protect the particular form of its  interactions. 
 The set-up spontaneously breaks Lorentz symmetry  by means of a time-like scalar field profile. By selecting a unitary gauge for the scalar, we show that 
  this theory describes the dynamics of 
a     spin two degree of freedom, whose
   equations of motion contain two time derivatives and  up to six
 spatial derivatives.
 We analytically determine asymptotically flat, 
spherically symmetric configurations,  showing that there exists
a  branch of  solutions  physically equivalent to spherically symmetric configurations
in General Relativity, also in the presence of matter fields.
\end{abstract}

\section{Introduction}

 Understanding the ultraviolet behaviour of Einstein gravity and its possible
 unitary completion at the Planck scale is one of the deepest open problems in high energy physics. 
   Stelle
 \cite{Stelle:1976gc} pointed out that    adding  specific covariant combinations of the Riemann tensor 
  to the Einstein-Hilbert action 
 makes 
 gravity renormalizable, at the price of introducing an Ostrogradsky ghost mode associated with higher
 order time derivatives in the field equations. Ho\v rava proposed 
 to eliminate the ghost by renouncing to Lorentz invariance, assigning  a special role to the time coordinate,  and considering combinations of 
 operators with up to six spatial derivatives of the metric, so to obtain
 a power counting renormalizable theory of
   gravity. Starting
 from the explicit original constructions \cite{Horava:2008ih,Horava:2009uw}, called Ho\v rava-Lifshitz gravity,  
  a large number of studies have explored this proposal in detail. 
Both the  projectable and non-projectable versions of the theory suffer from strong coupling problems and instabilities linked
 with the dynamics of a scalar mode  associated with the breaking of Lorentz invariance \cite{Charmousis:2009tc,Li:2009bg,Blas:2009yd}. The healthy extension of 
 the original theory \cite{Blas:2009qj} is free from such instabilities and belongs to a class of scenarios 
  called khronometric theories \cite{Blas:2010hb,Blas:2011ni}. Its spectrum -- besides the healthy scalar -- 
  includes  instantaneous interactions   
 which make subtle  the study of its physical consequences and the proof of  renormalizability beyond power
 counting \cite{Barvinsky:2015kil}.
  The scalar mode can be removed completely \cite{Horava:2010zj} by enriching the theory with auxiliary fields with an associated
  $U(1)$ symmetry,  so to make such  scalar
 a pure gauge mode. For the case of projectable
 Ho\v rava-Lifshitz gravity, {issues raised again this possibility  
   (see e.g. \cite{daSilva:2010bm, Huang:2010ay}), can be
   solved renouncing to the projectable condition: see e.g. \cite{Zhu:2011xe}}. There are many excellent
 reviews on all these topics, see e.g. \cite{Mukohyama:2010xz,Sotiriou:2010wn,Wang:2017brl}. 
 
 \smallskip

 In this work we build a covariant action for a new scalar-tensor theory  aimed to provide a  realisation of Ho\v rava
 proposal, with the specific property that it does not propagate scalar excitations around configurations where the
 scalar profile  satisfies a unitary gauge condition.  The theory
 propagates a single massless spin two tensor mode, that at low energies has the same dynamics as in GR, while at
 higher energies is characterised by modified dispersion relations as in Ho\v rava scenario.
 The absence of  scalar excitations is a welcome feature both from a theoretical and a
 phenomenological viewpoint, since it   automatically avoids
 strong coupling and instability issues in the scalar sector, as well as 
  constraints associated with
  possible
 gravitational and cosmological effects of so-far-unobserved light scalar modes.   
  
  A covariant version of  Ho\v rava-Lifshitz gravity is usually constructed by applying  a St\"uckelberg trick to the non-covariant version of  Ho\v rava theory: this approach  makes  manifest the dynamics of the scalar excitation associated   with the breaking of general covariance.  We take a different route, building our system directly in a covariant
  way by
  applying   
     techniques  that are currently being developed
  to investigate scalar-tensor theories, in particular the properties of disformal transformations \cite{Bekenstein:1992pj}.
  It is well known that a combination of disformal and conformal transformations (see
  e.g. \cite{Zumalacarregui:2013pma,Bettoni:2013diz,Crisostomi:2016tcp,Achour:2016rkg})
   can be used to 
  build classes of scalar-tensor theories characterized by higher order equations of motion, but that nevertheless
  do not propagate Ostrogradsky ghosts thanks to the presence of constraint conditions. Such examples belong
  to the class of covariant, degenerate higher order scalar-tensor theories \cite{Gleyzes:2014dya,Gleyzes:2014qga,Langlois:2015cwa,Crisostomi:2016czh,BenAchour:2016fzp}
    which propagate at most three
  degrees of freedom despite having equations of motion containing more than two time derivatives.
  
Here  we show that a suitable limit of a disformal transformation acting on a purely gravitational set-up leads to  a scalar-tensor system which spontaneously breaks Lorentz symmetry, selecting a time-like scalar profile. 
 The scalar-tensor interactions are invariant under a new symmetry acting on the metric. 
By selecting 
a  special foliation for the space-time --  which corresponds to a unitary gauge for the scalar field, and is  motivated by our pattern of spontaneous Lorentz
symmetry breaking --  we determine a set-up that does not propagate scalar degrees of freedom, but only a massless
spin two tensor mode, whose equations of motion contain two time  and at most six spatial derivatives.  Our system
can be seen as a  generalization of the cuscuton \cite{Afshordi:2006ad} and cuscuta-Galileon \cite{
Chagoya:2016inc,deRham:2016ged} theories. 
 The available
parameters
can be chosen such to reproduce GR results for the low energy, linearised propagation of tensor modes, while
the tensor dispersion relations are modified at high energies, potentially realising 
 Ho\v rava proposal of a power counting renormalizable  theory of gravity. Our system allows also to analytically
 study spherically symmetric configurations: we show that there are two  branches of asymptotically flat 
  solutions, one of which
 coincides with GR configurations, the other predicts subleading modifications 
 to GR results at large distances from a source.

\smallskip

  Our presentation proceeds as follows.  In Section \ref{sec-cons} we explain how we construct a theory with the  aforementioned properties.
 In  Section \ref{sec-dofs} we prove that the theory only propagates tensor modes 
 in a unitary gauge for the scalar, whose equations of motion are characterised by
  higher spatial derivatives. 
   In  Section \ref{sec-pheno} we discuss phenomenological consequences of modified  dispersion
   relations for linearised tensor perturbations, and the properties of spherically symmetric solutions. We conclude in Section \ref{sec-con}.

\section{The construction of the theory}\label{sec-cons}

We develop  a method for building     covariant
scalar-tensor  couplings that spontaneously break Lorentz invariance, and  whose properties
prevent   the propagation of scalar degrees of freedom in a unitary gauge. The resulting theory
propagates a single dynamical spin two mode, whose  dispersion relations are  modified at high energies with respect to Einstein gravity.

We start with an action $S_{gr}$ which is a combination
of covariant 
 Lagrangian densities involving only the metric tensor and its derivatives  
\bea \label{defasn}
{ S}_{gr}&=&\frac{M_{Pl}^2}{2}\,\int\,d^4 x\,\epsilon\, \sqrt{-g}\,\left[ {\cal L}_{R^{(1)}}
+   {\cal L}_{R^{(2)}}
+   {\cal L}_{R^{(3)}}
\right]\, ,
\eea
with $\epsilon$ a constant dimensionless parameter (that eventually  will be sent to zero) and 
\bea \label{deflr1}
 {\cal L}_{R^{(1)}}&=& {g_1}\,R\, ,
 \\
 M_{Pl}^2\,
  {\cal L}_{R^{(2)}}&=&
    {g_2} \,R^2 + \, g_3\, R_{\mu\nu} R^{\mu\nu} \, ,
  \\
   M_{Pl}^4\,
    {\cal L}_{R^{(3)}}&=&  {g_4} \,R^3 +
    g_5 \,R\,R_{\mu\nu} R^{\mu\nu}
    +g_6\,R_{\mu\rho} R^{\rho\sigma} R_{\sigma}^{\,\,\mu}+g_7\,R\,\Box \,R
    +g_8\,\left(\nabla_\mu R_{\nu \sigma}\right)^2\, ,
     \label{deflr3}
\eea
and $g_i$  are constants \footnote{If we were including also a cosmological constant
term, we would find that its disformal version would correspond to a cuscuton theory \cite{Afshordi:2006ad},
see \cite{Chagoya:2016inc}.}. Up to total derivatives, these 
are the most general parity preserving covariant combinations of Ricci curvature with up to six derivatives acting 
on the metric~\footnote{This set of Lagrangians was  also considered in \cite{Asorey:1996hz} to construct superrenormalizable models of quantum gravity, without considering disformal transformations.}. 
%
 We apply a specific disformal transformation  to the  action $S_{gr}$ of eq \eqref{defasn},  defined
in terms of a scalar field $\phi$ and the  parameter $\epsilon$ as
\be \label{gtmunu}
g_{\mu\nu}\to
g_{\mu\nu}-\frac{1}{\epsilon^2\,\Lambda^4}\,\partial_\mu \phi \partial_\nu \phi  \, ,
\ee
with inverse
\be
g^{\mu\nu}\,\to\,g^{\mu\nu}+\frac{1}{\epsilon^2\,\Lambda^4+X}\,\partial^\mu \phi\, \partial^\nu \phi \, ,
\ee
where
\be
X\,=\,- \partial^\mu \phi \partial_\mu \phi\, 
\ee
and $\Lambda$ is a mass scale that from now on we set to one: $\Lambda=1$. 
We are interested in  the `singular' limit of small (eventually vanishing) parameter  $\epsilon$: such limit leads to a theory with especially interesting
properties, at least in a unitary gauge for the scalar. \footnote{
This limit can have an interesting geometrical interpretation, first pointed out in \cite{Chagoya:2016inc},
in terms of an ultra-relativistic limit of a DBI-Galileon set-up. We will not develop this viewpoint further in this work. 
}
  
  The small $\epsilon$ limit of the disformal transformation, which we indicate with the symbol $\Rightarrow$, when  applied
  to the building blocks of the previous action gives
  \bea \label{gdis-df}
  g^{\mu\nu}&\Rightarrow&g_{dis}^{\mu\nu}\,=\,g^{\mu\nu}+\frac{1}{X}\,\partial^\mu \phi\, \partial^\nu \phi \, ,
\\
\sqrt{-g}&\Rightarrow&\sqrt{-g_{dis}}\,=\,\frac{1}{\epsilon}\sqrt{X}\,\sqrt{-g} \, ,
\\
R&\Rightarrow&R_{dis}\,=\,\frac{2 \left\langle \Phi ^2\right\rangle }{X^{2}}+\frac{\left[\Phi ^2\right]}{ {X}}+ {R }{}-\frac{2 \langle \Phi \rangle  ([\Phi ])}{X^{2}}-\frac{{[\Phi ]}^2}{{X}}+\frac{2 R^{\mu\nu} \partial_\mu\phi  \partial_\nu\phi }{{X}} \, ,
\\
R_{\mu\nu}&\Rightarrow&
R^{dis}_{\mu\nu} \,=\, R_{\mu \nu }+\frac{\partial ^{\alpha }\phi \Phi _{\alpha \mu }\Phi _{\beta \nu }\partial ^{\beta }\phi }{X^2}+\frac{R_{\mu \alpha \nu \beta }\partial ^{\alpha }\phi \partial ^{\beta }\phi }{X}+\frac{\Phi _{\alpha \mu } \Phi _{\nu }^{\alpha }}{X}-\frac{\Phi _{\mu \nu } \langle \Phi \rangle }{X^2}-\frac{\Phi _{\mu \nu } {[\Phi]}}{X}
\, ,
\nonumber\\
\label{rmunu-df}
  \eea
  where
   $\Phi$ is a matrix with components $\nabla^\mu \nabla_\nu \phi$, and
\begin{align}
[\Phi^n] & = \text{tr}(\Phi^n)\, , \\
\langle \Phi^n \rangle & = \partial \phi \cdot \Phi^n \cdot \partial\phi\, .
\end{align} 
In expressions \eqref{gdis-df}-\eqref{rmunu-df} we only write the  contributions that are leading in a small $\epsilon$
expansion, neglecting the higher order terms.  
The structure of the disformed action, which we call  $\tilde S_\phi$, can be obtained by  combining the results above: 
it is sufficient to substitute the disformed expressions for the inverse metric, the Ricci tensor, etc. contained in eqs  \eqref{gdis-df}-\eqref{rmunu-df} into the initial Lagrangians \eqref{deflr1}-\eqref{deflr3}. The resulting disformed system,
in the zero $\epsilon$ limit, 
is described by the scalar-tensor action
  $\tilde S_\phi$ 
 as
 \bea
{\tilde S}_\phi&=&\frac{M_{Pl}^2}{2}\,\int\,d^4 x\,\sqrt{-g}\,\left[ \tilde{\cal L}_{R^{(1)}}
+ \,  \tilde{\cal L}_{R^{(2)}}
+ \,  \tilde{\cal L}_{R^{(3)}}
\right]
\label{disfac1}
\\
&\equiv& \frac{M_{Pl}^2}{2}\,\int\,d^4x \,L_\phi \, .
 \eea
Notice that the factor of $\epsilon$ in eq \eqref{defasn} gets compensated by inverse powers of $\epsilon$ in the 
disformed quantities -- and terms with positive powers of $\epsilon$  vanish in the small $\epsilon$ limit. 
 The expression for  $ \tilde{\cal L}_{R^{(1)}}$ is relatively simple, but non-analytic since it contains
 a $\sqrt{X}$,  and reads
 \be \label{LR1a}
\tilde{\cal L}_{R^{(1)}}\,=\,g_1\,
{\sqrt{ \vphantom{X^2}X}}\left[R+\frac{{[\Phi ]}^2}{X}-\frac{{[\Phi^2 ]}}{X}
\right] \, .
\ee
 The   covariant  expressions for the $ \tilde{\cal L}_{R^{(2,\,3)}}$ are more complex, but -- as explained
 above -- can be straightforwardly obtained by combining
the disformed building blocks of  eqs \eqref{gdis-df}-\eqref{rmunu-df}. As an
 example, in 
Appendix 
\ref{appL2L3} we express the disformed Lagrangian $ \tilde{\cal L}_{R^{(2)}}$. Its covariant expression, and the one of
 $ \tilde{\cal L}_{R^{(3)}}$,
is again non-analytic. Their structure
makes them covariantized relatives of cuscuton and cuscuta-Galileon
theories \cite{Afshordi:2006ad,deRham:2016ged,Chagoya:2016inc},
which are known as not propagating scalar modes around a special foliation \cite{Bhattacharyya:2016mah,Gomes:2017tzd}. 

\smallskip

The resulting  theory has the following properties, which we will use in what follows:
\begin{itemize}
\item {\bf Symmetries:}
 Since the original theory, eq   \eqref{defasn},  is built only in terms of the metric, any transformation 
that leaves invariant the  combination built  in  \eqref{gtmunu}, corresponding to
the disformed metric,  will be a symmetry of the final theory $\tilde S_\phi$. One of such transformations
is
\bea
\phi&\to&\phi +\epsilon^2 \chi\, ,
\\
g_{\mu\nu}&\to& g_{\mu\nu}+\partial_\mu \chi \partial_\nu \phi 
+\partial_\mu \phi \partial_\nu \chi+\epsilon^2  \partial_\mu \chi \partial_\nu \chi\, ,
\eea
for an arbitrary scalar $\chi$. 
In the limit $\epsilon\to0$ we are considering, it reduces to a transformation that involves the metric only
\bea
g_{\mu\nu}&\to& g_{\mu\nu}+\partial_\mu \chi \partial_\nu \phi 
+\partial_\mu \phi \partial_\nu \chi \, .\label{newsym}
\eea
Consequently,  the transformation \eqref{newsym}
 is  a symmetry for the disformed action  $\tilde S_\phi$: it is 
 valid for any arbitrary scalar function $\chi$. A posteriori, this symmetry  
can motivate, and possibly protect,
 the specific structure of our scalar-tensor interactions that form $\tilde S_\phi$
 in eq \eqref{disfac1}.  
 We
 shall discuss in Section \ref{sec-dofs} its   consequences for characterising  the number
 of propagating degrees of freedom \footnote{This symmetry was noticed and used in related 
 contexts in \cite{Chagoya:2016inc,Filippini:2017kov,Chagoya:2017ojn}: it can be
 considered as a relative of the symmetries of DBI-Galileon system \cite{deRham:2010eu}, see also \cite{Chagoya:2016inc}
 for more details.}. 
 
 \item {\bf Preferred foliation of space-time:} Given  the non-analytic structure of our theory -- due to the presence of square roots and
 inverse powers of $X$, see eqs \eqref{gdis-df} --  the scalar necessarily  acquires  a time-like profile~\footnote{The
 scalar would select a space-like profile if we were writing the square root argument $\sqrt{-X}$ instead of $\sqrt{X}$ (see
   also \cite{Chagoya:2016inc}).
 }:   $\nabla_\mu \phi\, \nabla^\mu \phi <0$. This implies that
  Lorentz invariance is  spontaneously broken around
 any configuration solving the equations of motion. Given that
 Lorentz symmetry is broken, it is natural to select a preferred foliation for space-time
 corresponding to a {\it unitary gauge} for the scalar field, where constant time
 hypersurfaces coincide with constant scalar hypersurfaces. Such preferred
 foliation is special since around it a reduced number of degrees of freedom propagate, as we shall
 prove later. 
%
\item {\bf Constraint equation:} Besides symmetry  \eqref{newsym}, it is straightforward to
 explicitly check that
the `energy-momentum tensor' for this set-up, which we define as ($L_\phi$ given in~\eqref{disfac1})
\be \label{defemtsn}
T_{\mu\nu}\,=\,-\frac{2}{\sqrt{-g} }\,\frac{\delta { L}_\phi}{\delta g^{\mu \nu}}\, ,
\ee
automatically satisfies the relation
\be
T_{\mu \nu}\,\partial^\nu  \phi\,=\,0 \label{newb}
\ee
as a constraint, independently of the equations of motion: in a sense it is an analogue of the Bianchi identity satisfied by the Einstein
tensor in General Relativity.  We will make use of the covariant condition \eqref{newb} when studying the time
evolution of the ADM constraints in Section \ref{sec:adm}. 

\end{itemize}
\smallskip

The action ${\tilde S}_\phi$ has a peculiar feature: {\it it does not propagate tensor nor scalar degrees of freedom}, at
least in a unitary gauge where the scalar field $\phi$ depends only on time. 
This statement appears at first sight surprising:  the resulting action contains kinetically mixed
scalar and tensor  degrees of freedom, and moreover manifestly leads to equations of motion of order
higher than two -- hence potentially propagating Ostrogradsky ghosts. 

We shall develop a proof of  this statement in Section \ref{sec-dofs}: here we provide  some initial heuristic  arguments in its favour,
and discuss some of  its consequences in view of  a possible realisation of Ho\v{r}ava idea of a power-counting renormalizable
theory of gravity.  
 The fact that a theory described  by action ${\tilde S}_\phi$ does not propagate scalar modes, despite
manifestly containing a scalar field $\phi$, can be understood from the fact that the theory is built  from a disformal 
transformation of a theory $S_{gr}$ 
constructed only in terms of the metric tensor:
the final  action ${\tilde S}_\phi$ has the gauge
symmetry \eqref{newsym} that we can use to `gauge away' a propagating degree of freedom (analogously to the Maxwell
action for electromagnetism, where the $U(1)$ symmetry prevents the propagation of  the photon longitudinal mode, making it a pure gauge). The property that ${\tilde S}_\phi$ does not propagate tensors either, selecting a unitary gauge for the scalar
profile $\phi \,= \,\phi(t)$ is due to the fact that tensor fluctuations around any given background do not contain 
any time derivatives, but only {\it spatial} derivatives. This feature was noticed in \cite{Achour:2016rkg,deRham:2016wji,BenAchour:2016fzp} for the  particular case of $\tilde{\cal L}_{R^{(1)}}$,
 which is a special
case of quartic Horndeski. But  the
same property holds also for $\tilde{\cal L}_{R^{(2,\,3)}}$: in unitary gauge,  the action for the tensor mode contains   {\it spatial} derivatives only, that are not sufficient for making it 
 a propagating mode.

\bigskip

Since $\tilde{S}_{\phi}$ does not propagate any degree of freedom in unitary gauge, it might appear of little interest.  However,
we can add $\tilde{S}_{\phi}$ to the Einstein-Hilbert action, and write a complete system 
\bea \label{actcom}
S_T&=&S_{EH}+ \tilde{S}_{\phi}
\\
&=&\frac{M_{Pl}^2}{2}\,\int \sqrt{-g}\,R+ \tilde{S}_{\phi}\, .
\eea
The Einstein-Hilbert part provides a standard kinetic term for the tensor mode -- but the additional contribution of $ \tilde{S}_{\phi}$
 includes higher spatial derivatives to the action, and modify the tensor dispersion relations. The covariant 
 scalar-tensor theory $S_T$, once a unitary gauge $\phi=\phi(t)$ is selected, 
  contains
 spatial derivatives up to sixth order acting on the metric, and up to two time derivatives. Hence in a high energy regime -- where the effects of
 the higher derivatives in ${\tilde {\cal L}}_{R^{(3)}}$ become more and more relevant --  the theory becomes
 power counting renormalizable, providing  a specific realization of Ho\v rava proposal \cite{Horava:2009uw} \footnote{See also \cite{Afshordi:2009tt} for a connection between non-analytic
 scalar-tensor theories based on a cuscuton action and a low energy limit of  Ho\v rava-Lifshitz gravity.}.
  Interestingly, the presence of the
 Einstein-Hilbert term breaks the symmetry \eqref{newsym}, but  still the scalar-tensor system described
  by $S_T$ 
  does {\it not} propagate scalar excitations in the unitary gauge $\phi(t)$.

 
Since some of the contributions to our action are    obtained from a singular limit of a disformal transformation, our
scenario  might be related with the proposal
of mimetic gravity \cite{Chamseddine:2013kea}. A complete clarification
 of possible connections with this theory go beyond the scope of this paper:
on the other hand, we point out  that our final action for gravity is
the sum of the disformed action $\tilde S_\phi$ and the standard Einstein-Hilbert term, see eq \eqref{actcom}, which
is {\it not} obtained from a disformal transformation. Instead, an alternative realisation of a covariant renormalizable
theory of 
gravity which is more directly connected with mimetic gravity can be found in \cite{Cognola:2016gjy}.

\section{The propagating degrees of freedom}\label{sec-dofs}

We now investigate  the dynamics of propagating degrees of freedom for the action \eqref{actcom}. We first study the theory
in flat  space, and show  explicitly  that a residual symmetry remains that, in combination with diffeomorphisms, 
 prevents the propagation of scalar modes.
  We
 then make use of a covariant ADM formulation of the system  to show that, when a unitary gauge is
 selected for the scalar field,  the theory does not propagate
 scalars around any background, and the corresponding equations of motion act
  with higher spatial derivatives only on the tensor sector.

\subsection{A residual symmetry in    flat space-time removes the scalar excitation}

As mentioned in Section \ref{sec-cons}, the Einstein-Hilbert contribution to action \eqref{actcom} breaks the  
symmetry \eqref{newsym} obeyed by $\tilde S_\phi$: nevertheless a residual symmetry remains around
flat space, preventing
the propagation of scalar modes. We show this fact explicitly in this Section, discussing  an interesting 
 interplay  between
the symmetry  \eqref{newsym} and
 diffeomorphism invariance of the  system. 

 \smallskip
 
Computing the field equations   for the metric and scalar field associated with action \eqref{actcom}, one  finds they are satisfied for a
  flat space and for a time-like field profile
\be
 \label{scalan2}
\bar \phi\,=\,c_\mu x^\mu\, ,
\ee
for constant $c^\mu$ satisfying $c_\mu\,c_{\nu}\,
\eta^{\mu\nu} \,<\,0$. 
We can study the dynamics of fluctuations 
around a background metric and scalar
\bea
g_{\mu \nu}&=&\eta_{\mu\nu}+h_{\mu\nu}\, ,
\\
\phi&=&\bar \phi +\delta \phi\, .
\eea
We now exploit symmetry arguments to demonstrate  that, around flat space,  $\delta \phi$ is not dynamical since it is pure gauge.  The quadratic 
action for fluctuations associated with the total action $S_T$ of eq  \eqref{actcom}
 is made of two contributions. The Einstein-Hilbert
 part $S_{EH}$
  is invariant under infinitesimal
diffeomorphisms:
\be
h_{\mu\nu}\,\to\,h_{\mu\nu}+\partial_\mu \xi_\nu+\partial_\nu \xi_\mu\, ,
\ee
for arbitrary functions $\xi_\mu$. 
For what respect  $\tilde S_\phi$, we notice that the scalar field profile \eqref{scalan2} spontaneously breaks the 
diffeomorphism symmetry: under infinitesimal diffeos, the scalar fluctuation transforms non-linearly as
\be \label{scalL1tr}
\delta \phi \to  \delta \phi+\partial_\mu \bar \phi\,\xi^\mu \, .
\ee
On the other hand, we know that  $\tilde S_\phi$ is invariant under the transformation
 \eqref{newsym}:
\be \label{infsf1}
h_{\mu\nu}\to h_{\mu\nu}+\partial_\mu \bar \phi \,\partial_\nu \chi+\partial_\mu \chi \,\partial_\nu \bar \phi \, ,
\ee
for an arbitrary scalar function $\chi$.  This symmetry can be used 
 to `compensate' the spontaneous breaking of diffeomorphism 
  by  the scalar profile of eq \eqref{scalL1tr}. Indeed, exploiting
the transformation 
 \eqref{scalL1tr}, we can select a special unitary gauge choice (selecting appropriately a profile for  $\xi^\mu$)
 to  set to zero the scalar fluctuations, transferring  the scalar dof into the metric sector:
 \be
\bar \xi^\mu\,=\,\frac{\partial^\mu \bar \phi}{\bar X}\,\delta \phi\, ,
 \ee
with
\be
\bar X\,=\,-\partial_\rho \bar \phi \partial^\rho \bar \phi\, .
\ee
Making this choice of diffeomorphism parameter, we learn that under diffeomorphisms the field fluctuations transform as 
\bea
\delta \phi&\to&0\, ,
\\
h_{\mu\nu}&\to&h_{\mu\nu}+\partial_\mu \bar \xi_\nu+\partial_\nu  \bar \xi_\mu
\,=\,h_{\mu\nu}+\frac{1}{\bar X}\partial_\mu \bar \phi \,\partial_\nu \delta \phi
+\frac{1}{\bar X}\partial_\nu \bar \phi \,\partial_\mu \delta \phi\,  ,
 \label{hmetn2}
\eea
where we used the field profile of eq \eqref{scalan2}.
Such specific choice of diffeomorphism parameter $\xi^\mu$ does not affect the counting  of  
degrees of freedom in the 
 Einstein-Hilbert action (which
is invariant under diffeos) but it can affect $\tilde S_\phi$, where diffeos are spontaneously broken. 
  But
we can now use symmetry \eqref{infsf1} to compensate the contributions to the metric in eq \eqref{hmetn2}, by
applying transformation \eqref{infsf1} to eq \eqref{hmetn2} and 
 choosing 
\be
\chi\,=\,-\frac{\delta \phi}{\bar X} \, ,
\ee
so to make $h_{\mu\nu}$ invariant under a combination of diffeo and scalar transformations,
as desired.

Hence,   using the symmetries available around flat space
we learn that   scalar fluctuations
are pure gauge and do not propagate: diffeomorphisms and the scalar symmetry \eqref{newsym} combine together
to eliminate scalar fluctuations, even in the presence of a non-vanishing background scalar profile. 
In the next Section, we show with a more general method that the same remains
true around curved space-time, at least when selecting a unitary gauge for the scalar field.


\subsection{Counting the degrees of freedom in curved space-time in unitary gauge}\label{sec:adm}

We  now  consider the curved space-time case, where the  equations of motion
for the metric in our theory \eqref{actcom}
read
\be \label{eqco2}
G_{\mu\nu}\,=\,\frac{1}{M_{Pl}^2}\,T_{\mu\nu}\, ,
\ee
with $G_{\mu\nu}$ the Einstein tensor, and $T_{\mu\nu}$ the energy momentum tensor
 associated with $\tilde S_\phi$, as defined in eq \eqref{defemtsn}.

 As explained above, the non-analytic structure of our action spontaneously breaks Lorentz symmetry 
 by
 selecting a time-like profile for the scalar field. The broken Lorentz
 symmetry  motivates  the choice of a  special foliation for space-time, which we fix in terms of
  a {\it unitary gauge} where  the constant time hypersurfaces correspond to  constant
  scalar hypersurfaces. The covariant constraint \eqref{newb}, together with the equations of motion
  \eqref{eqco2}, lead to the relation 
  \begin{equation}
  \nabla^\mu \phi\,\left(M_{Pl}^2 G_{\mu \nu} - T_{\mu\nu} \right) \,=\,0 \, . \label{consug1}
  \end{equation}
  {The fact that the constraint \eqref{newb} holds identically -- independently of the equations of motion -- allows us to
  conclude that, in the unitary gauge, the relation above is  equivalent  to a contracted generalised Bianchi identity, ensuring  that the 
  Einstein tensor is preserved over time evolution normal to  constant time  
  hypersurfaces.  }

The unitary gauge for the scalar implies that $\phi$ depends only on time:
 without loss of generality we can assume
  the scalar Ansatz 
\be
\label{ugauge}
\phi\,=\,\phi_0\,t
\ee
for a constant $\phi_0$.

To identify the number of propagating degrees of freedom, it is convenient
to implement the covariant ADM description used in \cite{Langlois:2015skt}. 
The metric and 
its inverse are 
\begin{align}\label{eq:admmetric}
g_{\mu\nu} = \Bigg( \begin{array}{cc} -N^2 + \gamma_{ij} N^i N^j & \gamma_{ij} N^j  \\ 
\gamma_{ij} N^i & \gamma_{ij} \end{array}\Bigg)\, ,\hspace{1em}
g^{\mu\nu} = \frac{1}{N^2}\Bigg(  \begin{array}{cc} - 1 &  N^i  \\ 
N^j & N^2 \gamma^{ij} - N^i N^j \end{array}\Bigg)\, ,
\end{align}
where $N$ and $N^i$ are the lapse and shift functions, and $\gamma^{ij}$ is the 3d spatial metric. 
The unit normal to spatial
 hypersurfaces is \be
 n^\mu = N^{-1}(1, - N^i)  \,. \ee
  Following \cite{Langlois:2015skt}, we introduce
  the quantities
   $A_\mu = \partial_\mu \phi$,
and
\begin{equation}
A_* \equiv A_\mu n^\mu  = {N^{-1}}( A_0 - N^i A_i )\, .
\end{equation}
We now express our action  in terms of these quantities in  the unitary gauge \eqref{ugauge}. 
A straightforward calculation gives
\begin{align}
\nabla_0 A_{0} & =  N \dot A_* - A_* N^i N^j K_{ij} + N N^k D_k A_*\, , \nonumber \\
\nabla_i A_{0} & = - A_* N^j K_{ij} + N D_i A_*\, ,\nonumber \\
\nabla_i A_{j} & = - A_* K_{ij} \, , \nonumber \\
R& =\hat R + K_{ij}K^{ij} + K^2 - 2 N^{-1} (\partial_0 K - \mathcal L_{N^i} K + D_i D^i N) \, , \nonumber \\
R_{ij} & = \hat R_{ij} + K K_{ij} - 2  K_{ik}K^k_j - N^{-1} (\partial_0 K_{ij} - \mathcal L_{N^i} K_{ij} + D_iD_jN)\, .
\end{align}
In these expressions we indicate with $\hat R\,=\,{}^{(3)}R $ the Ricci  scalar (and with $\hat R_{ij}$ the Ricci tensor) calculated with the 3d spatial metric $\gamma_{ij}$, 
while {$D$ is the covariant derivative compatible with $\gamma$ and}
  $K_{ij}$ is the extrinsic curvature of spatial hypersurfaces:
  \be
  K_{ij}\,=\,\frac{1}{2 N} \left( \partial_0 \gamma_{ij}-D_i N_j-D_j N_i
  \right)\, .
  \ee
   We make use of  these  results 
in the definition  of the disformed action $\tilde S_\phi$ of eq \eqref{disfac1}.
 We find that 
   the disformed Lagrangians
in the limit $\epsilon\to0$ simplify enormously and 
 can be expressed as
\begin{align}
    \label{L1un}
\sqrt{- g}\,\tilde {\cal L}_{R^{(1)}}=& {g_1}\,\phi_0\,\sqrt{\gamma}\,\hat R \, ,
 \\
 M_{Pl}^2\,\sqrt{- g}\,
  \tilde {\cal L}_{R^{(2)}}=&\phi_0\,\sqrt{\gamma}\,\left(
    {g_2} \,\hat R^2 + \, g_3\, \hat R_{ij} \hat R^{ij } 
  \right) \, ,
      \label{L2un}
  \\
   M_{Pl}^4\,\sqrt{-g}\,
   \tilde  {\cal L}_{R^{(3)}}=& \phi_0\,\sqrt{\gamma}\,\left[  {g_4} \,\hat R^3 +
    g_5 \,\hat R\,\hat R_{ij} \,\hat R^{ij}
    +g_6\,\hat R^{\, i}_{j} \,\hat R^{\, j}_{m} \, \hat R^{\, m}_{i}+g_7\,\hat R\,\nabla^2  \, \hat R
    +g_8\,\left(\nabla_i \hat R_{j k}\right)^2
    \right] \, ,\nonumber\\
    \label{L3un}
\end{align}
where Latin indexes indicate spatial components. The previous expressions 
then combine into $\tilde S_\phi$ as in eq
\eqref{disfac1}, decomposed in 3+1 formalism in a unitary gauge for the scalar. Notice that, importantly, 
they  contain {\it only spatial derivatives} of the three dimensional metric, while
do not contain any time derivative in this gauge. Their
structure resembles the potential for the projectable version of Ho\v rava-Lifshitz gravity,
but  they {\it do not contain} the lapse function $N$ nor the shift $N_i$ in unitary gauge \eqref{ugauge}, thanks
to the presence of the overall  square roots $\sqrt{X}\,=\,\sqrt{-\phi_0^2\,g^{00}}\,=\,\phi_0\,N^{-1}$ in front of the Lagrangian densities.  The action $\tilde S_\phi$ can then be  linearly added to the standard ADM decomposition 
of the Einstein-Hilbert action $S_{EH}$.  

{
The  ADM constraint equations for the lapse and shift functions ($N$ and $N_i$) are not
modified with respect to GR, since the  $\tilde S_\phi$ contributions do not contain these quantities.
 This implies that these constraints still impose conditions preventing the propagation of would be degrees of freedom.
In GR,  
   these constraints are
preserved in time by virtue of the Bianchi identity, that in our scenario is substituted by 
the covariant 
 constraint \eqref{consug1}. A full Hamiltonian analysis of constraints in our theory is rather technical, and we
 relegate its discussion to Appendix \ref{app-ham}, at least for a special case of our system. There we show that the requirements of preserving constraints under time evolution
 add constraint conditions which prevent the propagation of the scalar degree of freedom.}

The equations of motion for the metric components $\gamma_{ij}$ only contain up  to two time derivatives from the Einstein-Hilbert
action, and up to six space derivatives from action $\tilde S_\phi$: the system  only propagates two transverse
traceless
tensor degrees of freedom in a unitary gauge, with modified dispersion relations. 

These results are all obtained  in the special foliation corresponding to a unitary gauge for the scalar, well  motivated 
by our symmetry breaking pattern leading to a time like scalar field, and 
where
calculations are much simpler and the results straightforward to obtain. We will briefly comment in the next Section on
 the possible behaviour of the system outside such gauge.
 \subsection{ On the geometrical interpretation of the disformal metric}
 In the limit $\epsilon\to0$, the disformed metric $g^{\mu\nu}_{dis}$ of eq.~\eqref{gdis-df}  effectively becomes a {\it spatial metric} 
  when evaluated in a unitary gauge.   Indeed, if 
 $\phi=\phi_0 t$ for a constant $\phi_0$, it is straightforward to verify that $g^{00}_{dis}=0$, $g^{0i}_{dis}=0$, and
 $g^{ij}_{dis} = g^{ij} - (g^{00})^{-1}g^{0i}g^{0j}$, so that only the spatial part of the disformal metric is non-vanishing.
 Furthermore, if $g$ is written in ADM form by using eq.~\eqref{eq:admmetric}, we can verify that $g^{ij}_{dis} = \gamma^{ij}$.
 The disformal metric $g^{\mu\nu}_{dis}$ can thus be seen as a projector onto the spatial slices of the ADM metric. With this
  result at hand, it is easy to understand why contractions of tensors such as $g^{\mu\nu}_{dis}R_{\mu\nu}^{dis}$ select only the spatial part of the relevant tensor, as shown in equations~\eqref{L1un}-\eqref{L3un}. 
  
  \smallskip
  
  The geometric structure described above is similar to what is used in the formulation of Newton-Cartan theory (see, e.g. \cite{Havas:1964zza}), a geometric reformulation of Newtonian gravity based on a contravariant symmetric tensor field, say $h_{NC}^{\mu\nu}$  with signature $(0,+,+,+)$, and  a covariant symmetric tensor field  $\tau^{NC}_{\mu\nu}$ with signature $(-,0,0,0)$ that can be written as
  $\tau^{NC}_{\mu\nu}=t_\mu t_\nu$ for a covariant vector field $t_\mu$. In addition there is a connection that defines a covariant derivative compatible with both $h_{NC}^{\mu\nu}$ and $t_\mu$, and the theory is written in a covariant way in terms of tensors constructed out of this connection. In our case, the role of $h_{NC}^{\mu\nu}$ is played by 
  $g^{\mu\nu}_{dis}$~\eqref{gdis-df}, and a natural candidate to be used in the definition of $t_\mu$ is $\partial_\mu\phi$. 
  On the other hand, in our framework these geometric structures are {\it not} the   fundamental  objects of our  theory, but
  quantities  derived  from the properties of our disformal transformations. In a sense,  our  approach could be seen
  as a way to obtain a Newton-Cartan theory starting from a covariant gravity system through  disformal relations.
   %
  %

 
\section{Phenomenological properties}\label{sec-pheno}

\subsection{Higher spatial derivatives and modified tensor dispersion relations}

Theories of gravity that break Lorentz symmetry, as khronometric  \cite{Blas:2010hb}, Einstein-Aether \cite{Jacobson:2000xp},  Ho\v rava-Lifshitz
 theories  
 need to satisfy strict theoretical and phenomenological constraints, 
 see
for example \cite{Blas:2014aca,Yunes:2016jcc} and the recent work \cite{Gumrukcuoglu:2017ijh},
 especially focussed on Ho\v rava-Lifshitz gravity after the GW170817 event.
An advantage of our set-up is that, in a unitary gauge for the background scalar field, only tensor modes propagate, with 
no scalar fluctuations: this fact automatically avoids various  unitarity and stability conditions, and helps to satisfy
  phenomenological constraints from 
  cosmology,  astrophysics, and solar system  deviations from GR. In this gauge, the  structure of the
  resulting theory -- see eqs \eqref{L1un}-\eqref{L3un} -- is very similar to 
the projectable version of Ho\v rava-Lifshitz gravity. At low energies, the higher spatial
 derivative contributions associated with \eqref{L2un}, \eqref{L3un}
  are suppressed, and the only deviations from GR are controlled by the contribution \eqref{L1un} (more on this later). 
At higher energies the contributions of the six spatial derivative Lagrangian  \eqref{L3un} becomes more and more  important: it is natural
to impose the Lifshitz scaling $z=3$, making gravity power counting renormalizable \cite{Horava:2009uw}. 
The projectable version of Ho\v rava-Lifshitz gravity
 has been shown to be renormalizable even beyond power counting \cite{Barvinsky:2015kil}, but it is affected by 
 instabilities or strong coupling
effects associated with the scalar excitation. Since the latter  is absent in our scenario in a unitary gauge for the scalar, it would be interesting
to investigate more in detail the full renormalizability properties of our set-up. 


Outside a unitary gauge, the equations of motion acquire higher order time derivatives, which
might indicate  that additional degrees   propagate, also  leading  to  instabilities: hence  the theory 
seems to be well behaved {\it only}  in the special frame associated with the time-dependent field profile \eqref{ugauge}. 
On the other hand, the recent work  \cite{DeFelice:2018ewo},
 using also arguments  developed in \cite{Blas:2009yd}, 
 gives some initial evidence that instantaneous modes  that appear
 to propagate outside the unitary gauge -- called shadowy modes -- can be made non-dynamical by choosing
 appropriate, physically  motivated boundary conditions \footnote{See also \cite{Maldacena:2011mk} for a proposal to eliminate the would-be
  ghost in conformal gravity by selecting appropriate boundary conditions for the fields.}. 
     A full understanding
  of this subject   is left for future work. 
  
We now briefly analyse the dynamics   
of linearised tensor fluctuations around flat space. They are defined  
   from the expression 
\be
d s^2\,=\,- d t^2+\left[ \delta_{ij}+\gamma_{ij}(t,\vec x)\right]\,d x^i d x^j \, ,
\ee
and are described by a  quadratic action for tensor modes
which is the same as 
in the projectable
version of Ho\v rava-Lifshitz theory:
\be
S^{(2)}_{tens}\,=\,\frac{M_{Pl}^2}{4}\int d t d^3 x\,\left[
\dot \gamma^2_{ij}-\left( \partial_m \gamma_{ij}\right)^2 - g_1\,\left( \partial_m \gamma_{ij}\right)^2
 + \frac{g_3}{2}\,\left( \partial_m  \partial_n\gamma_{ij}\right)^2 + \frac{g_8}{2}\,\left( \partial_m  \partial_n  \partial_k\gamma_{ij}\right)^2
\right]\, . \label{newco1}
\ee

This 
  action receives corrections 
  with respect to GR, 
   due to contributions  
    containing  up to six
   spatial derivatives acting on the tensor fluctuations,
   satisfying Ho\v rava conditions for a power counting renormalizable theory of gravity. 
     The GW170817 event severely constrains
the parameter $| g_1 |\le 10^{-15}$: from now, we set  $g_1=0$, and 
the  bounds discussed in \cite{Gumrukcuoglu:2017ijh} are avoided.
The remaining parameters $g_3$, $g_8$ in eq \eqref{newco1} can be constrained by absence
of observed gravitational  Cherenkov radiation \cite{Kiyota:2015dla,Yunes:2016jcc}, if 
they contribute in such a way to make the speed of gravity smaller than the speed of light for certain modes
with high momenta: their bounds are presently not very significant. 

\subsection{Spherically symmetric solutions}\label{sec-spsy}

The study of spherically symmetric solutions in Lorentz violating theories
has many 
  interesting and subtle implications, associated with the possibility to move faster than light and probe
   regions of space-time beyond  black hole horizons. Many studies of black holes in theories related
  with Ho\v rava gravity exist, see e.g. \cite{Barausse:2011pu,Blas:2011ni} and the review \cite{Barausse:2013nwa}. 

For our system, the study of spherically symmetric solutions is amenable of analytical treatment. Interestingly,
scalar-tensor systems  containing non-analytic functions of $X$, as square roots, have been considered in the
past for the study of spherically symmetric configurations in Horndeski theories,  for the possibility of evading
no-go theorems for the existence of black hole solutions. 
See e.g. the  papers \cite{Sotiriou:2013qea,Sotiriou:2014pfa,Babichev:2017guv} and the review
 \cite{Herdeiro:2015waa} for an extended discussion.  Our system fits well with
the class of theories where spherically symmetric configurations can be easily studied. 

We consider the following  spherically symmetric Ansatz for the metric and for the scalar in unitary gauge~\footnote{We verified that this Ansatz is sufficiently general and does not over-determine the metric: any more general Ansatz  for the fields does not lead to new independent equations of motion.}
\begin{align}
ds^2 &= - f(r) dt^2 + h(r)^{-1} dr^2 + 2 k(r) dt dr+ r^2 ( d\theta^2 +  \sin^2\theta)\, ,  \label{eq:metansatz}\\
\phi & = \phi_0 \,t\,.   \label{eq:sfansatz}
\end{align}
For simplicity, we focus on the case $g_1\,=\,g_7 \,=\, g_8 \,=\, 0$, and
we set the Planck mass $M_{Pl}\,=\,1$. The equation of motion for the metric
component $k(r)$ reduces to an algebraic equation,
%
%
%
\begin{equation} \label{algcon}
k\, \left( f - f h^{-1} - k^2 + r f' \right) = 0\, .
\end{equation}
This condition bifurcates  the solutions in two disconnected branches,
\begin{eqnarray}
&&{\text{First branch:}}\,\hskip1cmk^2 =  f - f h^{-1} + r f'  \, ,
\\
&&{\text{Second branch:}}\,\hskip0.7cm k =0 \, .
\end{eqnarray}

%
%
%
We discuss in turn some features of each branch.

\smallskip
\noindent
$\bullet$  {\bf First Branch:}
In the first branch the equations of motion lead to the following solution
\begin{align}
h(r) & = 1 \, , \nonumber \\
f(r) & = 1 - \frac{2 M}{r} \, ,  \label{stecom}\\
k(r) & = \sqrt{\frac{2 M}{r}}\, ,\nonumber 
\end{align}
that does not   depend on any of the parameters $g_i$ controlling the action. The physical properties
of this configuration are identical to the ones of a Schwarzschild black hole written
 in Painleve-Gullstrand coordinates~\footnote{See \cite{Sotiriou:2009gy,Greenwald:2011ca} for a discussion of the use of these coordinates
 in the context of the   projectable
version of  Ho\v rava-Lifshitz gravity.}, 
making this solution a 
stealth configuration. 

 Indeed, in GR  a metric with components \eqref{stecom} can be easily 
recasted in the form of Schwarzschild black hole by 
 a change of coordinates $d t\to dt+k(r)\,dr/f(r)$, which removes the cross component of the metric.
  This implies that the metric \eqref{stecom} has the same physical features of the
  Schwarzschild solution in GR, {expressed} in a particular coordinate system.
  
  \smallskip
  
  More is actually true for this branch of configurations. Even in presence of matter, as long as there is no direct coupling between the scalar and 
matter fields described by an action $S_m$,
 the spherically symmetric solutions for the complete scalar-tensor system plus matter, described by the action
   $ S_T+S_m$ are identically to the ones of GR plus
 matter fields, whose equations of motion are associated with the action $S_{EH}+S_m$. We prove this fact in Appendix 
 \ref{app-branch}.  This property implies that, within this branch, the physical properties of 
  spherically symmetric solutions are the same 
 as GR configurations, including  solutions describing gravitationally bound  compact objects, 
 as spherically symmetric stars,
 etc.
 Differences between GR and our  theory  probably arise when one considers  departures from spherical symmetry, and
 studies for example axially symmetric and rotating configurations.

\smallskip
\noindent
$\bullet$  {\bf Second Branch:} The second branch of solutions corresponds to the choice $k(r)=0$ in
equation \eqref{algcon}.  In this case, the metric is automatically diagonal.  
Also in this situation an analytic, asymptotically flat solution exists, given by
 (recall that we are choosing 
for simplicity $g_1\,=\,g_7 \,=\, g_8 \,=\, 0$)

\begin{figure}\centering
	\includegraphics[width=0.6\textwidth]{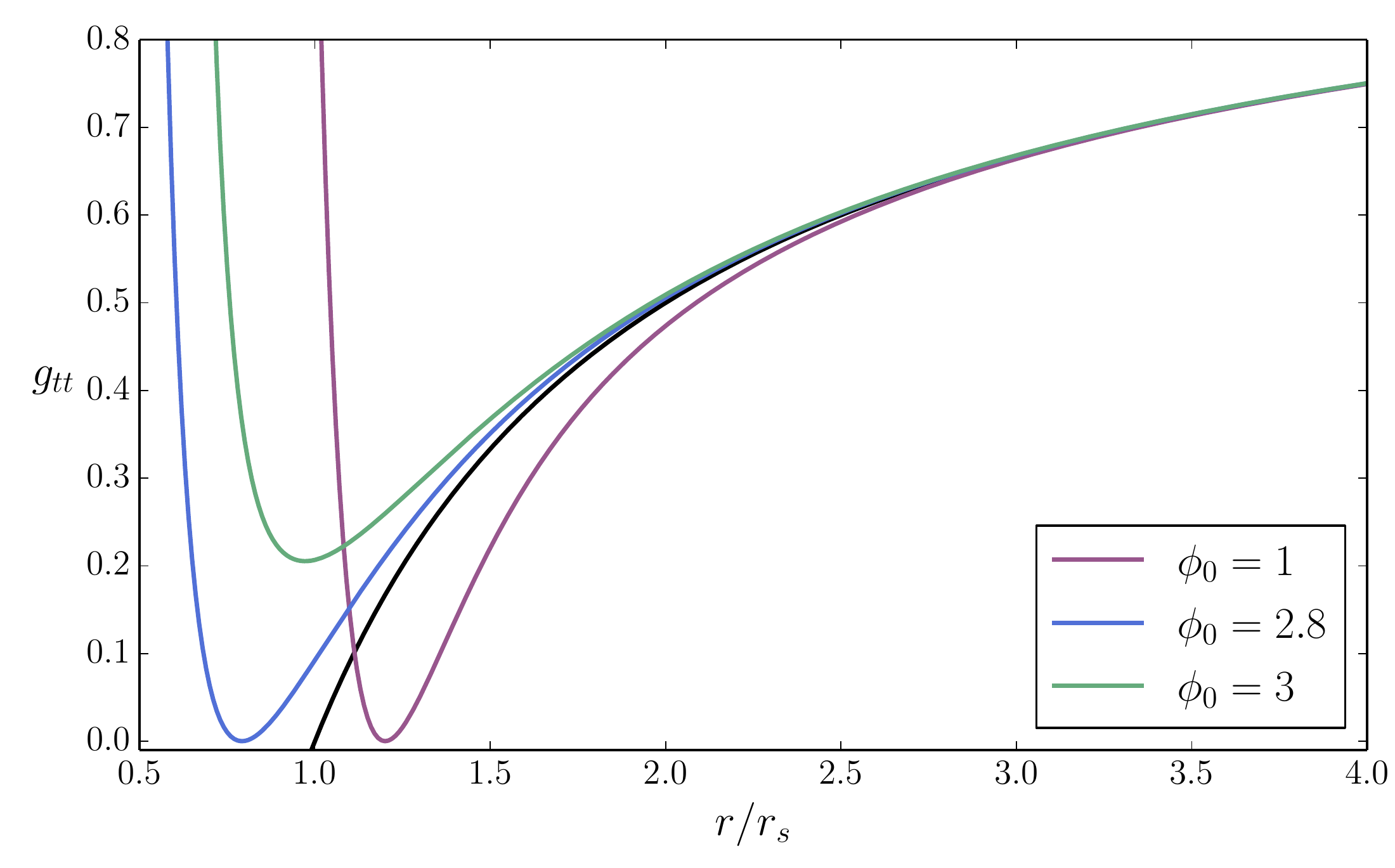}\caption{\it $g_{tt}$ component of the metric in the second branch of solutions, eq.~(\ref{eq:gttns}), for $g_1= 0, g_3=-2, g_5=1, M=1$. The black solid line shows the Schwarzschild solution. Although all the solutions are asymptotically flat, large deviations from Schwarzschild can exist near the horizon. The points where $g_{tt}=0$ correspond to divergences of the Ricci scalar.}\label{fig:nonsch}
\end{figure}

\begin{align}
h(r)  = & 1 -\frac{2M}{r}\, , \label{eq:grrns}\\
f(r)  = & \left[ \sqrt{1-\frac{2 M}{r}} 
 -g_3 \phi _0\frac{M^2(M + r)+ 2 r^2 (M - r + \sqrt{r (r-2M)}) }{10 M^2 r^3} 
 + g_5 \phi _0\frac{6 M^2 }{r^6}  \right. \nonumber \\
& \left. + g_6 \phi _0 \frac{1050 M^6+7 M^5 r+5 M^4 r^2+4 M^2 r^3 (r + M)+ 8 r^5(M - r + \sqrt{r (r-2M)} )}{154 M^4 r^6} \right]^2 \,  \label{eq:gttns}\, ,
\end{align}
where an integration constant has been used in order to ensure asymptotic flatness: 
 $f\to 1$ as $r\to \infty$. For large $r$, $f(r)$ behaves as
\begin{equation}
f = 1  - \frac{2M}{r}  
 - 
 \frac{g_3\, \phi_0 M^2 }{4 r^4} + \cdots\, .
\end{equation}
Thus, this solution describes asymptotic corrections to the Schwarzschild metric that start at order $1/r^4$
at large distances. The profile of $f(r)$ is shown in Fig.~\ref{fig:nonsch} for fixed $g_i$ and varying $\phi_0$, with values
 justified below.
In general, conditions on the $g_i$ have to be imposed in order to keep the geometry regular. First of all, the square roots in~\eqref{eq:gttns} introduce complex terms in the metric that are also reflected in the curvature invariants. These terms are removed by the condition
\begin{equation}
-20 g_6 \phi _0+77 M^2 g_3 \phi _0+385 M^4 
 =0\, .
\end{equation}
Another condition comes from examining the Ricci scalar. Using the equations of motion, $R$ can be written as
\begin{equation}
R = \frac{3 M^2 \left[r^3 g_3+15 r (8 g_5+9 g_6)-12 M (22 g_5+25 g_6)\right] \phi _0}{r^9 \sqrt{f}}\, .\label{eq:ricciscalar}
\end{equation}
Thus, it generally diverges when $f=0$. Analytically finding the zeros of $f$ is complicated. Numerically we find that, given a set of $g_i$'s, the
position where $f=0$ can be controlled with $\phi_0$. For $\phi_0>0$, this position can be pushed to $r< 2M$ if at least one of the $g_i$'s is negative.

 An alternative argument is the following: since $f$ is always positive, $f'$ must change sign when $f= 0$. Asymptotically, $f'>0$. At $r=2M$, $f'$ 
 takes the value
 \begin{equation}
  -\frac{3 \left[271 M^2 g_3 \phi _0+60 g_5 \phi _0+1155 M^4 \right]\left[215 M^2 g_3 \phi _0+48 g_5 \phi _0+1155 M^4 \right] }{163840 M^9}\, .
\end{equation}
If $f'>0 $ at $r=2M$, this suggests (although does not guarantee) that $f \neq 0$ for every $r>2M$. 
Requiring that one of the factors above is negative can be seen as a condition on $\phi_0$, and requiring that $\phi_0$ is positive implies that 
at least one $g_i$ is negative. For the values of $M$ and $g_i$ used in Fig.~\ref{fig:nonsch}, the first factor becomes negative when $\phi_0\gtrsim 2.39$, and the second
factor becomes negative when $\phi_0\gtrsim 3.02$. This agrees with the behaviour observed in Fig.~\ref{fig:nonsch}.

  \section{Conclusions}\label{sec-con}

In this work we introduced a covariant scalar-tensor theory of gravity aimed to provide
a new realisation of Ho\v rava proposal for a power-counting renormalizable theory of gravity. 
Our construction 
 spontaneously breaks Lorentz symmetry through a non-vanishing scalar profile, and it 
has the special feature of not propagating scalar excitations around a foliation
of space-time corresponding to a unitary gauge for the scalar field. We proved this fact using a covariant
ADM method, and  we also discussed  the result in terms of the symmetries of the system.
  The  dynamical 
degree of freedom in this gauge is a spin two tensor mode with modified dispersion relations:
its equations of motion contain two time and up to six spatial derivatives, potentially realising
Ho\v rava's idea. Our scalar-tensor interactions are invariant under  a specific symmetry acting
on the metric, which might protect their structure, and are related with cuscuton
\cite{Afshordi:2006ad} and cuscuta-Galileon \cite{Chagoya:2016inc,deRham:2016ged} theories. 
We analytically determined spherically symmetric solutions for our theory, showing that there
exists a branch of stealth solutions physically equivalent to General Relativity configurations, also
in the presence of matter fields.

We  conclude  stressing again  parallelisms and differences with other frameworks. In  our
scenario  Lorentz symmetry is {\it spontaneously broken}, instead of explicitly broken as in the original construction of \cite{Horava:2008ih,Horava:2009uw}. 
We start from covariant actions, whose structure (as the presence of $\sqrt{X}$ coefficients) leads to a time dependent scalar profile, and spontaneous breaking of Lorentz invariance, even around
flat space. Interestingly, the structure of scalar-tensor couplings, derived from the application of our specific choice disformal transformation, is such that
 scalar excitations do not propagate in an unitary gauge for the scalar. Hence, we do not need to worry about the  behaviour and possible strong coupling issues
 associated with an extra mode, which require  delicate treatments or extra ingredients in the original theory (see e.g. \cite{Blas:2009yd,
 Horava:2010zj,daSilva:2010bm,Huang:2010ay,Zhu:2011xe,Cognola:2016gjy}).

Much  work is left for the future. It would be important to study whether this system can lead to a theory
of  gravity renormalizable beyond power counting: the theory shares many similarities with the projectable
version of Ho\v rava gravity -- which is known to be renormalizable \cite{Barvinsky:2015kil} --  with the advantage of being free of dangerous scalar excitations.
{ Additional work is needed in order to fully understand the  stability of the structure of our construction, and whether
the symmetry \eqref{newsym} can prevent the danger that 
 additional Lorentz-violating operators can be generated at the quantum level.} 
 At the phenomenological level, it would be interesting to study in general terms possible consequences of our construction  for cosmology, and 
  to include more general parity violating interactions
as well, along the lines of \cite{Crisostomi:2017ugk}. We hope to report soon on these
issues.
 
\subsection*{Acknowledgments}

It is a pleasure to thank
Marco Crisostomi, Emir Gumrukcuoglu,  Maria Mylova and Sergey Sibiryakov for interesting discussions, an
anonymous referee for many suggestions to improve our work, 
and Guillem Dom\`enech for comments on the manuscript. 
We are partially supported by the STFC grant ST/P00055X/1.

\begin{appendix}

\section{The expression for the disformed Lagrangians}\label{appL2L3}
The Lagrangian density $\tilde {\cal L}_{R^{(2)}}$ is given by

\begin{align}
\tilde {\cal L}_{R^{(2)}} = & g_2 \,
{\sqrt{ \vphantom{X^2}X}}\left[\frac{2 \left\langle \Phi ^2\right\rangle }{X^{2}}+\frac{\left[\Phi ^2\right]}{ {X}}+ {R }{}-\frac{2 \langle \Phi \rangle  ([\Phi ])}{X^{2}}-\frac{{[\Phi ]}^2}{{X}}+\frac{2 R^{\mu\nu} \partial_\mu\phi  \partial_\mu\phi }{{X}} \right]^2 \nonumber
\\
& + g_3\, \sqrt{X}\,\bigg[
 \left[R^2\right] +\frac{2 \langle R^2\rangle }{{X}}+R_{{ \sigma\rho}}\left(\frac{2 \Phi ^{{\mu \rho }}\Phi _\mu ^\sigma}{{X}} +\frac{2 \langle R\rangle^{{\sigma \rho }} }{{X}}  -\frac{2 \Phi ^{{\sigma\rho}} {[\Phi ]}}{{X}} \right) +\frac{\left[\Phi ^4\right]}{X^{2}} +\frac{{\langle R\rangle }^2}{X^{2}} \nonumber  \\
&+\frac{4\langle R_{{\nu \rho }} \Phi _\mu ^\sigma\Phi _\sigma^\rho \rangle}{X^{2}} +\langle R\rangle _{{\gamma \delta }} \left(\frac{\langle R\rangle^{{\gamma \delta }} }{X^{2}} +\frac{2 \Phi ^{{\sigma \delta }}\Phi _\sigma^\gamma }{X^{2}}  -\frac{2 \Phi ^{{\gamma \delta }} {[\Phi ]}}{X^{2}} \right)-\frac{2 R_{{\gamma\delta}}\Phi ^{{\gamma \delta}} \langle \Phi \rangle }{X^{2}}  -\frac{2 \left[\Phi ^3\right] {[\Phi ]}}{X^{2}}\nonumber \\ 
&-\frac{4 \langle R_{{\nu \sigma}}\Phi _\mu ^\sigma\rangle  {[\Phi ]}}{X^{2}} +\frac{2 \langle R_{{\sigma \rho }} \Phi _\mu ^\sigma\Phi _\nu ^\rho \rangle }{X^{2}}+\frac{{[\Phi ]}^2 \left[\Phi ^2\right]}{X^{2}}  -\frac{6 \left\langle \Phi ^3\right\rangle  {[\Phi ]}}{X^{3}}  +\frac{2 \langle  \langle R\rangle_{{\gamma \delta }}  \Phi _\mu ^\gamma \Phi _\nu ^\delta \rangle }{X^{3}}+2\frac{\left[\Phi ^4\right] + \langle \Phi ^4\rangle }{X^{3}}  \nonumber \\
&+\frac{2 \left\langle \Phi ^2\right\rangle  {[\Phi ]}^2}{X^{3}}+\frac{2 \left\langle \Phi ^2\right\rangle  {\langle R\rangle }}{X^{3}} -\frac{2 \langle R\rangle _{{ \gamma \delta }} \Phi ^{{\gamma \delta }} \langle \Phi \rangle }{X^{3}}-\frac{2 \left[\Phi ^3\right] \langle \Phi \rangle }{X^{3}} +\frac{2 \left[\Phi ^2\right] {[\Phi ]} \langle \Phi \rangle }{X^{3}}-\frac{2 {[\Phi ]} {\langle R\rangle } \langle \Phi \rangle }{X^{3}} \nonumber \\ 
&-\frac{2 \left\langle \Phi ^2\right\rangle  \langle \Phi \rangle }{X^{4}}-\frac{2 \left\langle \Phi ^2\right\rangle  {[\Phi ]} \langle \Phi \rangle }{X^{4}} +\frac{2 \left\langle \Phi ^2\right\rangle ^2}{X^{4}}+\frac{\left[\Phi ^2\right] \langle \Phi \rangle^2  }{X^{4}}+\frac{{[\Phi ]}^2 \langle \Phi \rangle^2 }{X^{4}}
\bigg ]\, .
\end{align}

Here,  $\Phi$ is a matrix with components $\nabla^\mu \nabla_\nu \phi$, and
\begin{align}
X & = - \partial^\mu \phi \partial_\mu \phi\, , \\
[\Phi^n] & = \text{tr}(\Phi^n)\, , \\
\langle \Phi^n \rangle & = \partial \phi \cdot \Phi^n \cdot \partial\phi\, ,  \\
\langle R \rangle_{\mu\nu} & = \partial^\alpha \phi \partial^\beta \phi R_{\alpha \mu\beta\nu }\, ,
\end{align} 
and where indices are explicit, the free indices inside $\langle \ \rangle$ are contracted with two first derivatives of the scalar field.

 Although the expressions for the disformal Lagrangians can be quite cumbersome, obtaining them is not a complicated task.
The idea is simple: compute the Christoffel symbols for the disformal metric, and use these symbols to construct the geometric quantities of interest. The new  Christoffel symbols $(\Gamma^{dis})^{\alpha}_{\mu\nu}$ are those of the original metric plus contributions from
the scalar field. We can use this fact to simplify the calculations and express the new geometric quantities in terms of the old ones plus terms coming from the difference between the Christoffel symbols of $g_{\mu\nu}$ and of the disformal metric. For example, the transformed Riemann tensor is\cite{Zumalacarregui:2013pma}
\begin{equation}
(R^{dis})^\alpha{}_{\beta\mu\nu} = R^\alpha{}_{\beta\mu\nu} + 2 \nabla_{[\mu}\mathcal K^\alpha{}_{\nu]\beta} + 2 \mathcal K^\alpha{}_{\gamma[\mu}\mathcal K^\gamma{}_{\nu]\beta}\, ,
\end{equation}
where, for the  disformal transformation~\eqref{gtmunu}, $\mathcal K^\alpha{}_{\mu\nu}= (\Gamma^{dis})^{\alpha}_{\mu\nu}-\Gamma^\alpha_{\mu\nu}$ takes the form
\begin{equation}
\mathcal K^\alpha{}_{\mu\nu}=- \frac{\partial^\alpha\phi \nabla_\mu\partial_\nu\phi}{\epsilon^2-\partial^\beta\phi \partial_\beta\phi} \, .
\end{equation}
With the help of symbolic computation packages such as \emph{xAct}\cite{xact}, it is straightforward to compute the contractions of
the Riemann tensor required to obtain the disformal Lagrangians, being careful to transform also the covariant derivatives when necesseary, e.g. when computing the transformation of $\Box R$.

\section{Hamiltonian analysis}
\label{app-ham}

In this Appendix we develop a Hamiltonian analysis of a special case of our system, showing that it propagates only a spin-two
degree of freedom in unitary gauge. 

We focus on the Einstein-Hilbert action supplemented with the singular disformal transformation of the Ricci scalar, associated with Lagrangian
\eqref{LR1a}.  When expressed in unitary gauge and making use of a 3+1 ADM formalism, the action reads
 \begin{equation}\label{ac:RRdisf}
 S(\gamma, N, N^i) = \int dt L= \int dt \, d^3 x \sqrt{\gamma} N \left( \hat R + K_{ij}K^{ij} + K^2  \right) + g_1 \phi_0 \sqrt{\gamma} \hat R\, 
 \end{equation}
 where we have ignored a boundary term from the Einstein-Hilbert part of the action. The dynamics
  (time derivatives) of the metric are contained only in the extrinsic curvature.

  In order to write  the Hamiltonian for \eqref{ac:RRdisf}, we have to compute the canonical momenta conjugated to $\gamma$, $N$, and $N^i$. Since no time derivatives of the lapse and shift functions appear in the action, the only non-vanishing momentum is
  \begin{equation}
  \tilde P^{ij} = \frac{\delta L}{\delta \dot \gamma_{ij}} = \sqrt{q} (K^{ij} - K \gamma^{ij})\, .
  \end{equation}
  The time derivatives of the metric can be written in terms of the momenta, mod. the undetermined functions $N$ and $N^i$. Thus the Hamiltonian reads
\begin{equation}
\label{hams1}
H = \int d^3 x N\left( -\sqrt{\gamma} \hat R  + \frac{1}{\sqrt{\gamma}} \left( \tilde{P}^{ij}\tilde{P}_{ij} - \frac{1}{2} \tilde{P}^2 \right) \right) - 2 N^i \gamma_{ij} D_k \tilde{P}^{jk} -  g_1 \phi_0 \sqrt{\gamma} \hat R \, .
\end{equation}

\smallskip




The lapse and shift functions play the role of Lagrange multipliers enforcing the constraints
\begin{align}
P_N & =   -\sqrt{\gamma} \hat R  + \frac{1}{\sqrt{\gamma}} \left( \tilde{P}^{ij}\tilde{P}_{ij} - \frac{1}{2} \tilde{P}^2 \right) \approx 0 \, ,\\
P_i & =  - 2 \gamma_{ij} D_k \tilde{P}^{jk} \approx 0 \, ,
\end{align}
which define the constraint functions $C(N) = \int d^3 x N P_N$ and $C(\vec N) = \int d^3 x N^i P_i$. Notice that these constraints are the same as in GR.

The Poisson bracket between two real functions, say $f$ and $g$, in phase space is given by
\begin{equation} \label{eq:pbracket}
\left\{ f,g  \right\} = \int d^3 x \frac{\delta f}{\delta \gamma_{ij}} \frac{\delta g}{\delta \tilde{P}^{ij}} - \frac{\delta f}{\delta \tilde P^{ij}} \frac{\delta g}{\delta \gamma_{ij}}
\end{equation}
It is possible to verify that the constraint functions satisfy
\begin{align}
\{ C(\vec N), C(\vec M)   \} & = C(\mathcal L_N M^i) \, , \\
\{ C(\vec N), C(M)  \}  & = C(\mathcal L_N M)\, ,\\
\{  C(N), C(M)  \} & = C(\vec K)\, , 
\end{align}
where $K^i = \gamma^{ij} (N \partial_j M - M \partial_j N)$. This implies that the constraint functions form a first class set. 

In contrast to GR, the total Hamiltonian is not only a sum of these first class constraints:  it contains additional
pieces proportional to $g_i$ -- in the specific case we are considering here, the last term in eq.~ \eqref{hams1}. Thus, the 
fact that our first class constraints are the same as in GR does not automatically guarantee that they are preserved in time: additional 
constraints associated with this requirement can prevent the propagation of scalar degrees of freedom.

An interesting, general approach for directly discussing such conditions for preventing the propagation of scalar modes in unitary gauge has been developed in\cite{Iyonaga:2018vnu}.
  Their  starting point is the following general structure for an action  in ADM variables, where the extrinsic curvature is replaced by an auxiliary variable $Q_{ij}$ by means
 of Lagrange multipliers:
\begin{equation}
S = \int dt d^3 x \, N \sqrt{\gamma} \left[ L(t, N, \gamma_{ij}, \hat R_{ij}, Q_{ij}) + v^{ij} (Q_{ij} - K_{ij})\right]\, .\label{eq:admaction}
\end{equation}
Our action \eqref{ac:RRdisf} can be easily expressed as above.  The canonical momenta of
\eqref{eq:admaction} define a set of primary constraints, whose preservation in
time defines a set of secondary constraints. The preservation in time of the secondary constraints leads to a linear  system of equations that can usually be solved in terms of   the Lagrange multipliers that enforce the primary constraints in the total Hamiltonian. However, if this is not possible due to degeneracy of the system, one then needs to impose further tertiary constraints, which reduce the number of propagating degrees of freedom.  In the case of an action of the form eq. \eqref{eq:admaction}, such degeneracy condition reads (see \cite{Iyonaga:2018vnu} for full details):

\begin{equation}
\Delta \equiv (N L)_{NN} - N L^{ij}_{QN}(L^{-1}_{QQ})_{ij,kl}L^{kl}_{QN} = 0\,,
\label{eq:degcon}
\end{equation}

\noindent
 where suffixes as $L_N$ etc mean derivatives of the Lagrangian along the fields in the index. It is straightforward to show that our Lagrangian satisfies this condition. 
 Indeed, 
  in our case  the combination $N L$ is composed of two parts, one that depends linearly in $N$, and another one that does not depend on $N$, so that $L_{NN} = 0$. We are then left to show that the second term in $\Delta$ vanishes: the additional terms in our Lagrangian do not depend on the extrinsic curvature and therefore do not depend either on the auxiliary variably $Q$,  on the other hand, the GR part does not depend on $N$, thus $L_{NQ} = 0$. In conclusion, the action  \eqref{ac:RRdisf} 
 leads to the degeneracy condition   
 $\Delta = 0$, which implies the existence of a tertiary constraints forbidding the propagation of a scalar excitation in unitary gauge: our theory then propagates
 only a spin-2 mode when the scalar satisfies an unitary gauge.
 
 A more direct way to reach the same result is to make use of the findings around \eqref{consug1}, where we have
 seen that
 the identity~\eqref{newb} translates into the conservation of the constraints. Thus, the model we are considering here has the same
number of initial degrees of freedom, and the same number and type of constraints as GR, therefore it propagates only
two degrees of freedom.


\section{Spherically symmetric solutions with matter fields}
\label{app-branch}

In this Appendix we reconsider the first branch of spherically symmetric solutions discussed
in Section \ref{sec-spsy}, but we extend our arguments to consider an arbitrary matter
sector minimally coupled to gravity. The action we consider is
\be \label{ext-ac}
S_{(g_{\mu\nu} + \phi+matter)}
\,=\,\frac{M_{Pl}^2}{2}\,\int d^4 x\,\sqrt{-g}\,R\,+\,\tilde S_\phi+\int d^4 x\,\sqrt{-g}\,{\cal L}_m \, ,
\ee
where $\tilde S_\phi$ is given in equation \eqref{disfac1}, and ${\cal L}_m$ a matter Lagrangian. We have
learned in Section  \ref{sec-spsy} that in absence of matter the theory admits a branch of 
spherically symmetric, stealth configurations which are physically equivalent  to the solutions
of Einstein gravity. Here we show that the same remains true also in presence of matter, as long as
 there are no direct couplings between matter and scalar field, as in eq \eqref{ext-ac}.  In other
 words, we show that there is a branch of configurations where the spherically symmetric solutions
 of \eqref{ext-ac} coincide with the ones of the reduced action
 \be \label{red-ac}
S_{(g_{\mu\nu} +matter)}
\,=\,\frac{M_{Pl}^2}{2}\,\int d^4 x\,\sqrt{-g}\,R\,+\,\int d^4 x\,\sqrt{-g}\,{\cal L}_m \, .
\ee
This feature is not unique to our scalar-tensor theory, and has been  recently proved to
be shared with vector-tensor solutions \cite{Chagoya:2017ojn} in vector-tensor theories of gravity first
introduced in \cite{Tasinato:2014eka,Heisenberg:2014rta,Tasinato:2014mia}.  

The argument goes as follow (see  \cite{Chagoya:2017ojn} for more details in a very similar set-up). 
One starts from a general static spherically symmetric
 Ansatz for the metric \footnote{We checked that our arguments remain valid also for a more general spherically
 symmetric time dependent metric Ansatz.}, 
 \begin{align}
		ds^2 &= - f(r) dt^2 + h(r)^{-1} dr^2 + 2 k(r) dt dr+ r^2 ( d\theta^2 +  \sin^2\theta)\, , 
	\end{align}
	and a general spherically symmetric Ansatz for the scalar, $\phi(t,r)$. We derive the equations
	of motion for the metric components, and for the scalar field associated
	with action \eqref{ext-ac}. We select a unitary gauge 
	$\phi(t,r)\,=\,\phi_0 t$ for the scalar profile, and choose an Ansatz $h(r)\,=\,1$ for one of the metric
	components: this profile for $h$ characterizes the stealth branch of solutions in vacuum, see eq \eqref{stecom}.
	{We find that the scalar field equation automatically vanishes, and
		the  remaining equations for the metric components $f(r)$, $k(r)$ are completely independent of the 
		scalar, and are the same equations one would derive from the action \eqref{red-ac} by selecting the Ansatz $h(r)=1$.}
	This implies that the spherically symmetric solutions are the same as in GR, as we wish to demonstrate. Within
	GR there is no preferred foliation, and one can choose a gauge different from $h=1$ by changing coordinates:
nevertheless the physical properties of the solutions are the same in any coordinate system.

\end{appendix}


\begin{thebibliography}{95}
{
	\bibitem{Stelle:1976gc}
	K.~S.~Stelle,
	``Renormalization of Higher Derivative Quantum Gravity,''
	Phys.\ Rev.\ D {\bf 16} (1977) 953.
	doi:10.1103/PhysRevD.16.953
	
	
	\bibitem{Horava:2008ih}
	P.~Horava,
	``Membranes at Quantum Criticality,''
	JHEP {\bf 0903} (2009) 020
	doi:10.1088/1126-6708/2009/03/020
	[arXiv:0812.4287 [hep-th]].
	
	\bibitem{Horava:2009uw}
	P.~Horava,
	``Quantum Gravity at a Lifshitz Point,''
	Phys.\ Rev.\ D {\bf 79} (2009) 084008
	doi:10.1103/PhysRevD.79.084008
	[arXiv:0901.3775 [hep-th]].
	
	\bibitem{Blas:2009yd}
	D.~Blas, O.~Pujolas and S.~Sibiryakov,
	``On the Extra Mode and Inconsistency of Horava Gravity,''
	JHEP {\bf 0910} (2009) 029
	doi:10.1088/1126-6708/2009/10/029
	[arXiv:0906.3046 [hep-th]].
	
	\bibitem{Charmousis:2009tc}
	C.~Charmousis, G.~Niz, A.~Padilla and P.~M.~Saffin,
	``Strong coupling in Horava gravity,''
	JHEP {\bf 0908} (2009) 070
	doi:10.1088/1126-6708/2009/08/070
	[arXiv:0905.2579 [hep-th]].
	
	\bibitem{Li:2009bg}
	M.~Li and Y.~Pang,
	``A Trouble with Horava-Lifshitz Gravity,''
	JHEP {\bf 0908} (2009) 015
	doi:10.1088/1126-6708/2009/08/015
	[arXiv:0905.2751 [hep-th]].
	
	\bibitem{Blas:2009qj}
	D.~Blas, O.~Pujolas and S.~Sibiryakov,
	``Consistent Extension of Horava Gravity,''
	Phys.\ Rev.\ Lett.\  {\bf 104} (2010) 181302
	doi:10.1103/PhysRevLett.104.181302
	[arXiv:0909.3525 [hep-th]].
	
	\bibitem{Blas:2010hb}
	D.~Blas, O.~Pujolas and S.~Sibiryakov,
	``Models of non-relativistic quantum gravity: The Good, the bad and the healthy,''
	JHEP {\bf 1104} (2011) 018
	doi:10.1007/JHEP04(2011)018
	[arXiv:1007.3503 [hep-th]].
	
	\bibitem{Blas:2011ni}
	D.~Blas and S.~Sibiryakov,
	``Horava gravity versus thermodynamics: The Black hole case,''
	Phys.\ Rev.\ D {\bf 84} (2011) 124043
	doi:10.1103/PhysRevD.84.124043
	[arXiv:1110.2195 [hep-th]].
	

	
	\bibitem{Barvinsky:2015kil}
	A.~O.~Barvinsky, D.~Blas, M.~Herrero-Valea, S.~M.~Sibiryakov and C.~F.~Steinwachs,
	``Renormalization of Horava gravity,''
	Phys.\ Rev.\ D {\bf 93} (2016) no.6,  064022
	doi:10.1103/PhysRevD.93.064022
	[arXiv:1512.02250 [hep-th]].
	
	
	\bibitem{Horava:2010zj}
	P.~Horava and C.~M.~Melby-Thompson,
	``General Covariance in Quantum Gravity at a Lifshitz Point,''
	Phys.\ Rev.\ D {\bf 82} (2010) 064027
	doi:10.1103/PhysRevD.82.064027
	[arXiv:1007.2410 [hep-th]].
	
	\bibitem{daSilva:2010bm}
	A.~M.~da Silva,
	``An Alternative Approach for General Covariant Horava-Lifshitz Gravity and Matter Coupling,''
	Class.\ Quant.\ Grav.\  {\bf 28} (2011) 055011
	doi:10.1088/0264-9381/28/5/055011
	[arXiv:1009.4885 [hep-th]].
	
	\bibitem{Huang:2010ay}
	Y.~Huang and A.~Wang,
	``Stability, ghost, and strong coupling in nonrelativistic general covariant theory of gravity with $\lambda \not=1$,''
	Phys.\ Rev.\ D {\bf 83} (2011) 104012
	doi:10.1103/PhysRevD.83.104012
	[arXiv:1011.0739 [hep-th]].
	
\bibitem{Zhu:2011xe}
  T.~Zhu, Q.~Wu, A.~Wang and F.~W.~Shu,
  Phys.\ Rev.\ D {\bf 84} (2011) 101502
  doi:10.1103/PhysRevD.84.101502
  [arXiv:1108.1237 [hep-th]];
  T.~Zhu, F.~W.~Shu, Q.~Wu and A.~Wang,
  Phys.\ Rev.\ D {\bf 85} (2012) 044053
  doi:10.1103/PhysRevD.85.044053
  [arXiv:1110.5106 [hep-th]];
  K.~Lin, S.~Mukohyama, A.~Wang and T.~Zhu,
  Phys.\ Rev.\ D {\bf 89} (2014) no.8,  084022
  doi:10.1103/PhysRevD.89.084022
  [arXiv:1310.6666 [hep-ph]].




	\bibitem{Sotiriou:2010wn}
	T.~P.~Sotiriou,
	``Horava-Lifshitz gravity: a status report,''
	J.\ Phys.\ Conf.\ Ser.\  {\bf 283} (2011) 012034
	doi:10.1088/1742-6596/283/1/012034
	[arXiv:1010.3218 [hep-th]].
	
	\bibitem{Mukohyama:2010xz}
	S.~Mukohyama,
	``Horava-Lifshitz Cosmology: A Review,''
	Class.\ Quant.\ Grav.\  {\bf 27} (2010) 223101
	doi:10.1088/0264-9381/27/22/223101
	[arXiv:1007.5199 [hep-th]].
	
	\bibitem{Wang:2017brl}
	A.~Wang,
	``Horava gravity at a Lifshitz point: A progress report,''
	Int.\ J.\ Mod.\ Phys.\ D {\bf 26} (2017) no.07,  1730014
	doi:10.1142/S0218271817300142
	[arXiv:1701.06087 [gr-qc]].
	
	\bibitem{Bekenstein:1992pj}
	J.~D.~Bekenstein,
	``The Relation between physical and gravitational geometry,''
	Phys.\ Rev.\ D {\bf 48} (1993) 3641
	doi:10.1103/PhysRevD.48.3641
	[gr-qc/9211017].
	
	\bibitem{Bettoni:2013diz}
	D.~Bettoni and S.~Liberati,
	``Disformal invariance of second order scalar-tensor theories: Framing the Horndeski action,''
	Phys.\ Rev.\ D {\bf 88} (2013) 084020
	doi:10.1103/PhysRevD.88.084020
	[arXiv:1306.6724 [gr-qc]].
	
	\bibitem{Zumalacarregui:2013pma}
	M.~Zumalac\'{a}rregui and J.~Garc\'{i}a-Bellido,
	``Transforming gravity: from derivative couplings to matter to second-order scalar-tensor theories beyond the Horndeski Lagrangian,''
	Phys.\ Rev.\ D {\bf 89} (2014) 064046
	doi:10.1103/PhysRevD.89.064046
	[arXiv:1308.4685 [gr-qc]].
	
		\bibitem{Crisostomi:2016tcp}
	M.~Crisostomi, M.~Hull, K.~Koyama and G.~Tasinato,
	JCAP {\bf 1603} (2016) no.03,  038
	doi:10.1088/1475-7516/2016/03/038
	[arXiv:1601.04658 [hep-th]].
	
	\bibitem{Achour:2016rkg}
	J.~Ben Achour, D.~Langlois and K.~Noui,
	``Degenerate higher order scalar-tensor theories beyond Horndeski and disformal transformations,''
	Phys.\ Rev.\ D {\bf 93} (2016) no.12,  124005
	doi:10.1103/PhysRevD.93.124005
	[arXiv:1602.08398 [gr-qc]].

	
	 
	\bibitem{Gleyzes:2014dya}
	J.~Gleyzes, D.~Langlois, F.~Piazza and F.~Vernizzi,
	``Healthy theories beyond Horndeski,''
	Phys.\ Rev.\ Lett.\  {\bf 114} (2015) no.21,  211101
	doi:10.1103/PhysRevLett.114.211101
	[arXiv:1404.6495 [hep-th]].
	
	\bibitem{Gleyzes:2014qga}
	J.~Gleyzes, D.~Langlois, F.~Piazza and F.~Vernizzi,
	``Exploring gravitational theories beyond Horndeski,''
	JCAP {\bf 1502} (2015) 018
	doi:10.1088/1475-7516/2015/02/018
	[arXiv:1408.1952 [astro-ph.CO]].
	
	\bibitem{Langlois:2015cwa}
	D.~Langlois and K.~Noui,
	``Degenerate higher derivative theories beyond Horndeski: evading the Ostrogradski instability,''
	JCAP {\bf 1602} (2016) no.02,  034
	doi:10.1088/1475-7516/2016/02/034
	[arXiv:1510.06930 [gr-qc]].
	
	\bibitem{Crisostomi:2016czh}
	M.~Crisostomi, K.~Koyama and G.~Tasinato,
	``Extended Scalar-Tensor Theories of Gravity,''
	JCAP {\bf 1604} (2016) no.04,  044
	doi:10.1088/1475-7516/2016/04/044
	[arXiv:1602.03119 [hep-th]].
	
	\bibitem{BenAchour:2016fzp}
	J.~Ben Achour, M.~Crisostomi, K.~Koyama, D.~Langlois, K.~Noui and G.~Tasinato,
	``Degenerate higher order scalar-tensor theories beyond Horndeski up to cubic order,''
	JHEP {\bf 1612} (2016) 100
	doi:10.1007/JHEP12(2016)100
	[arXiv:1608.08135 [hep-th]].
	
	\bibitem{Afshordi:2006ad}
	N.~Afshordi, D.~J.~H.~Chung and G.~Geshnizjani,
	``Cuscuton: A Causal Field Theory with an Infinite Speed of Sound,''
	Phys.\ Rev.\ D {\bf 75} (2007) 083513
	doi:10.1103/PhysRevD.75.083513
	[hep-th/0609150].
	
	
	\bibitem{Chagoya:2016inc}
	J.~Chagoya and G.~Tasinato,
	``A geometrical approach to degenerate scalar-tensor theories,''
	JHEP {\bf 1702} (2017) 113
	doi:10.1007/JHEP02(2017)113
	[arXiv:1610.07980 [hep-th]].
	
	\bibitem{deRham:2016ged}
	C.~de Rham and H.~Motohashi,
	``Caustics for Spherical Waves,''
	Phys.\ Rev.\ D {\bf 95} (2017) no.6,  064008
	doi:10.1103/PhysRevD.95.064008
	[arXiv:1611.05038 [hep-th]].
	

\bibitem{Asorey:1996hz}
  M.~Asorey, J.~L.~Lopez and I.~L.~Shapiro,
  Int.\ J.\ Mod.\ Phys.\ A {\bf 12} (1997) 5711
  doi:10.1142/S0217751X97002991
  [hep-th/9610006];
  L.~Modesto and I.~L.~Shapiro,
  Phys.\ Lett.\ B {\bf 755} (2016) 279
  doi:10.1016/j.physletb.2016.02.021
  [arXiv:1512.07600 [hep-th]].


	\bibitem{Bhattacharyya:2016mah}
	J.~Bhattacharyya, A.~Coates, M.~Colombo, A.~E.~Gumrukcuoglu and T.~P.~Sotiriou,
	``Revisiting the cuscuton as a Lorentz-violating gravity theory,''
	Phys.\ Rev.\ D {\bf 97} (2018) no.6,  064020
	doi:10.1103/PhysRevD.97.064020
	[arXiv:1612.01824 [hep-th]].
	
	\bibitem{Gomes:2017tzd}
	H.~Gomes and D.~C.~Guariento,
	``Hamiltonian analysis of the cuscuton,''
	Phys.\ Rev.\ D {\bf 95} (2017) no.10,  104049
	doi:10.1103/PhysRevD.95.104049
	[arXiv:1703.08226 [gr-qc]].
	
	\bibitem{Filippini:2017kov}
	F.~Filippini and G.~Tasinato,
	``An exact solution for a rotating black hole in modified gravity,''
	JCAP {\bf 1801} (2018) no.01,  033
	doi:10.1088/1475-7516/2018/01/033
	[arXiv:1709.02147 [hep-th]].
	
	\bibitem{Chagoya:2017ojn}
	J.~Chagoya and G.~Tasinato,
	``Stealth configurations in vector-tensor theories of gravity,''
	JCAP {\bf 1801} (2018) no.01,  046
	doi:10.1088/1475-7516/2018/01/046
	[arXiv:1707.07951 [hep-th]].
	
\bibitem{deRham:2010eu}
  C.~de Rham and A.~J.~Tolley,
  ``DBI and the Galileon reunited,''
  JCAP {\bf 1005} (2010) 015
  doi:10.1088/1475-7516/2010/05/015
  [arXiv:1003.5917 [hep-th]].

\bibitem{Chamseddine:2013kea}
  A.~H.~Chamseddine and V.~Mukhanov,
  ``Mimetic Dark Matter,''
  JHEP {\bf 1311} (2013) 135
  doi:10.1007/JHEP11(2013)135
  [arXiv:1308.5410 [astro-ph.CO]].

\bibitem{Cognola:2016gjy}
  S.~Nojiri and S.~D.~Odintsov,
  Phys.\ Rev.\ D {\bf 81} (2010) 043001
  doi:10.1103/PhysRevD.81.043001
  [arXiv:0905.4213 [hep-th]];
  G.~Cognola, R.~Myrzakulov, L.~Sebastiani, S.~Vagnozzi and S.~Zerbini,
  Class.\ Quant.\ Grav.\  {\bf 33} (2016) no.22,  225014
  doi:10.1088/0264-9381/33/22/225014
  [arXiv:1601.00102 [gr-qc]].

	
	\bibitem{deRham:2016wji}
	C.~de Rham and A.~Matas,
	``Ostrogradsky in Theories with Multiple Fields,''
	JCAP {\bf 1606} (2016) no.06,  041
	doi:10.1088/1475-7516/2016/06/041
	[arXiv:1604.08638 [hep-th]].
	
	\bibitem{Afshordi:2009tt}
	N.~Afshordi,
	``Cuscuton and low energy limit of Horava-Lifshitz gravity,''
	Phys.\ Rev.\ D {\bf 80} (2009) 081502
	doi:10.1103/PhysRevD.80.081502
	[arXiv:0907.5201 [hep-th]].
	
	\bibitem{Langlois:2015skt}
	D.~Langlois and K.~Noui,
	``Hamiltonian analysis of higher derivative scalar-tensor theories,''
	JCAP {\bf 1607} (2016) no.07,  016
	doi:10.1088/1475-7516/2016/07/016
	[arXiv:1512.06820 [gr-qc]].
	
	\bibitem{Frittelli:1996nj}
	S.~Frittelli,
	``Note on the propagation of the constraints in standard (3+1) general relativity,''
	Phys.\ Rev.\ D {\bf 55} (1997) 5992.
	doi:10.1103/PhysRevD.55.5992
	
	\bibitem{Havas:1964zza}
	P.~HAVAS,
	Rev.\ Mod.\ Phys.\  {\bf 36} (1964) 938.
	doi:10.1103/RevModPhys.36.938
	
	
	\bibitem{Jacobson:2000xp}
	T.~Jacobson and D.~Mattingly,
	``Gravity with a dynamical preferred frame,''
	Phys.\ Rev.\ D {\bf 64} (2001) 024028
	doi:10.1103/PhysRevD.64.024028
	[gr-qc/0007031].
	
	\bibitem{Blas:2014aca}
	D.~Blas and E.~Lim,
	``Phenomenology of theories of gravity without Lorentz invariance: the preferred frame case,''
	Int.\ J.\ Mod.\ Phys.\ D {\bf 23} (2015) 1443009
	doi:10.1142/S0218271814430093
	[arXiv:1412.4828 [gr-qc]].
	
	\bibitem{Yunes:2016jcc}
	N.~Yunes, K.~Yagi and F.~Pretorius,
	``Theoretical Physics Implications of the Binary Black-Hole Mergers GW150914 and GW151226,''
	Phys.\ Rev.\ D {\bf 94} (2016) no.8,  084002
	doi:10.1103/PhysRevD.94.084002
	[arXiv:1603.08955 [gr-qc]].
	
	\bibitem{Gumrukcuoglu:2017ijh}
	A.~Emir G{\"u}mr{\"u}k{\c c}{\"u}o{\v g}lu, M.~Saravani and T.~P.~Sotiriou,
	``Ho{\v r}ava gravity after GW170817,''
	Phys.\ Rev.\ D {\bf 97} (2018) no.2,  024032
	doi:10.1103/PhysRevD.97.024032
	[arXiv:1711.08845 [gr-qc]].
	
	\bibitem{Kiyota:2015dla}
	S.~Kiyota and K.~Yamamoto,
	``Constraint on modified dispersion relations for gravitational waves from gravitational Cherenkov radiation,''
	Phys.\ Rev.\ D {\bf 92} (2015) no.10,  104036
	doi:10.1103/PhysRevD.92.104036
	[arXiv:1509.00610 [gr-qc]].
	
	\bibitem{DeFelice:2018ewo}
	A.~De Felice, D.~Langlois, S.~Mukohyama, K.~Noui and A.~Wang,
	``"Shadowy" modes in Higher-Order Scalar-Tensor theories,''
	arXiv:1803.06241 [hep-th].
	
	\bibitem{Barausse:2011pu}
	E.~Barausse, T.~Jacobson and T.~P.~Sotiriou,
	``Black holes in Einstein-aether and Horava-Lifshitz gravity,''
	Phys.\ Rev.\ D {\bf 83} (2011) 124043
	doi:10.1103/PhysRevD.83.124043
	[arXiv:1104.2889 [gr-qc]].
	
	\bibitem{Barausse:2013nwa}
	E.~Barausse and T.~P.~Sotiriou,
	``Black holes in Lorentz-violating gravity theories,''
	Class.\ Quant.\ Grav.\  {\bf 30} (2013) 244010
	doi:10.1088/0264-9381/30/24/244010
	[arXiv:1307.3359 [gr-qc]].
	
	\bibitem{Maldacena:2011mk}
	J.~Maldacena,
	``Einstein Gravity from Conformal Gravity,''
	arXiv:1105.5632 [hep-th].
	
	\bibitem{Sotiriou:2013qea}
	T.~P.~Sotiriou and S.~Y.~Zhou,
	``Black hole hair in generalized scalar-tensor gravity,''
	Phys.\ Rev.\ Lett.\  {\bf 112} (2014) 251102
	doi:10.1103/PhysRevLett.112.251102
	[arXiv:1312.3622 [gr-qc]].
	\bibitem{Sotiriou:2014pfa}
	T.~P.~Sotiriou and S.~Y.~Zhou,
	``Black hole hair in generalized scalar-tensor gravity: An explicit example,''
	Phys.\ Rev.\ D {\bf 90} (2014) 124063
	doi:10.1103/PhysRevD.90.124063
	[arXiv:1408.1698 [gr-qc]].
	
	\bibitem{Babichev:2017guv}
	E.~Babichev, C.~Charmousis and A.~Leh\'{e}bel,
	``Asymptotically flat black holes in Horndeski theory and beyond,''
	JCAP {\bf 1704} (2017) no.04,  027
	doi:10.1088/1475-7516/2017/04/027
	[arXiv:1702.01938 [gr-qc]].
	
	
	\bibitem{Herdeiro:2015waa}
	C.~A.~R.~Herdeiro and E.~Radu,
	``Asymptotically flat black holes with scalar hair: a review,''
	Int.\ J.\ Mod.\ Phys.\ D {\bf 24} (2015) no.09,  1542014
	doi:10.1142/S0218271815420146
	[arXiv:1504.08209 [gr-qc]].
	
\bibitem{Sotiriou:2009gy}
  T.~P.~Sotiriou, M.~Visser and S.~Weinfurtner,
  ``Phenomenologically viable Lorentz-violating quantum gravity,''
  Phys.\ Rev.\ Lett.\  {\bf 102} (2009) 251601
  doi:10.1103/PhysRevLett.102.251601
  [arXiv:0904.4464 [hep-th]].


\bibitem{Greenwald:2011ca}
  J.~Greenwald, J.~Lenells, J.~X.~Lu, V.~H.~Satheeshkumar and A.~Wang,
  ``Black holes and global structures of spherical spacetimes in Horava-Lifshitz theory,''
  Phys.\ Rev.\ D {\bf 84} (2011) 084040
  doi:10.1103/PhysRevD.84.084040
  [arXiv:1105.4259 [hep-th]].


	\bibitem{Tasinato:2014eka}
	  G.~Tasinato,
	  ``Cosmic Acceleration from Abelian Symmetry Breaking,''
	  JHEP {\bf 1404} (2014) 067
	  doi:10.1007/JHEP04(2014)067
	  [arXiv:1402.6450 [hep-th]]

	
	\bibitem{Heisenberg:2014rta}
	  L.~Heisenberg,
	  ``Generalization of the Proca Action,''
	  JCAP {\bf 1405} (2014) 015
	  doi:10.1088/1475-7516/2014/05/015
	  [arXiv:1402.7026 [hep-th]].
	
	\bibitem{Tasinato:2014mia}
  G.~Tasinato,
  ``A small cosmological constant from Abelian symmetry breaking,''
  Class.\ Quant.\ Grav.\  {\bf 31} (2014) 225004
  doi:10.1088/0264-9381/31/22/225004
  [arXiv:1404.4883 [hep-th]].
  
  
\bibitem{Crisostomi:2017ugk}
  M.~Crisostomi, K.~Noui, C.~Charmousis and D.~Langlois,
  ``Beyond Lovelock gravity: Higher derivative metric theories,''
  Phys.\ Rev.\ D {\bf 97} (2018) no.4,  044034
  doi:10.1103/PhysRevD.97.044034
  [arXiv:1710.04531 [hep-th]].

\bibitem{xact}
http://www.xact.es

\bibitem{Iyonaga:2018vnu}
  A.~Iyonaga, K.~Takahashi and T.~Kobayashi,
  ``Extended Cuscuton: Formulation,''
  arXiv:1809.10935 [gr-qc].

%
%
%
%
%
%
%
%
%
%
%
%
%
%
%
%
%
%
%


}

\end{thebibliography}
\end{document}